\documentclass{aa}  

\usepackage{graphicx}
\usepackage{txfonts}
\usepackage{subcaption}
\usepackage{hyperref}
\usepackage{cleveref}
\begin{document} 

  \title{One star, two star, red star, blue star: an updated planetary nebula central star distance catalogue from \emph{Gaia} EDR3
  }
  \titlerunning{An updated planetary nebula central star distance catalogue from \emph{Gaia} EDR3}

  \author{N. Chornay \and N. A. Walton}

  \institute{Institute of Astronomy, University of Cambridge, Madingley Road, Cambridge, CB3 0HA, United Kingdom\\
              \email{njc89@ast.cam.ac.uk}, \email{naw@ast.cam.ac.uk}}

  \date{September 2021}

 
  \abstract
  {Planetary nebulae (PNe) are a brief but important phase of stellar evolution. The study of Galactic PNe has historically been hampered by uncertain distances, but the parallaxes of PN central stars (CSPNe) measured by \textit{Gaia} are improving the situation.}
  {\textit{Gaia}'s Early Data Release 3 (EDR3) offers higher astrometric precision and greater completeness compared to previous releases. Taking advantage of these improvements requires that the CSPNe in the catalogue be accurately identified.}
  {We applied our automated technique based on the likelihood ratio method to cross-match known PNe with sources in \textit{Gaia} EDR3, using an empirically derived position and colour distribution to score candidate matches.}
  {We present a catalogue of over 2000 sources in \textit{Gaia} EDR3 that our method has identified as
  likely
  CSPNe or compact nebula detections. We show how the more precise parallaxes of these sources compare to previous PN statistical distances and introduce an approach to combining them to produce tighter distance constraints. We also discuss \textit{Gaia}'s handling of close companions and bright nebulae.}
  {\textit{Gaia} is unlocking new avenues for the study of PNe. The catalogue presented here will remain valid for the upcoming \textit{Gaia} Data Release 3 (DR3) and thus provide a valuable resource for years to come.}


  \maketitle
%

\section{Introduction}

Planetary nebulae (PNe) are a brief but enigmatic stage in the evolution of low- and intermediate-mass stars towards the ends of their lives. Central stars of PNe (CSPNe) are some of the rarest objects observed by \textit{Gaia} \citep{gaiamission}, a European Space Agency (ESA) mission conducting precision astrometry and photometry for nearly two billion sources in the Milky Way and beyond. \textit{Gaia} shows great potential for the study of PNe: parallaxes inform distance estimates and in turn CSPN absolute magnitudes and evolutionary phase; proper motions add more dimensions to dynamics and relate PNe to their parent stellar populations, constraining ages and masses; and repeated photometric observations enable the study of variability and binarity.

Unlocking \textit{Gaia}'s potential requires a complete and accurate cross-match of Gaia sources with CSPNe. \citet[][hereinafter CW20]{chornay2020cspn} used an automated approach to generate an extensive catalogue of candidate CSPNe in \textit{Gaia} Data Release 2 \citep[DR2,][]{gaiadr2}.
The nature of \textit{Gaia}'s iterative release process necessitates updates to catalogues based on \textit{Gaia} data, as the set of sources and their identifications changes between releases \citep{torra2021gaiaedr3sourcelist}. Moreover, when the \textit{Gaia} data is itself used for cross-matching, better data translates into improved accuracy.
In this work we present a new catalogue of candidate CSPNe and compact PNe in \textit{Gaia} Early Data Release 3 \citep[EDR3,][]{gaiacollaboration2021edr3survey} based on the same methodology as CW20. With a completely new \textit{Gaia} catalogue not expected for some time, this will provide a valuable resource for the study of Galactic 
PNe for years to come.

\section{Methods} \label{sect:methods}

\subsection{Input catalogues}

Our PN positions come from the Hong Kong/AAO/Strasbourg H$\alpha$ (HASH) PN catalogue \citep[][as of 5 July 2021]{hashpn}, which now contains 2670 "true" PNe (those with spectroscopic confirmation; an 8\% increase since CW20) and 3800 PNe in total (including likely and possible identifications). We saw in CW20 that three very extended PNe had CSPNe more than 60\arcsec\ away from the HASH positions, outside of our search radius. To address this we adopt positions for Sh 2-188 from \citet{weidmannCSPNe} and for FP J1824-0319 and FP J0905-3033 from \citet{mashpn}. 
\textit{Gaia} EDR3 is likewise larger than DR2. The longer survey duration (12 more months, for 34 months in total), along with improved calibration, results in higher completeness (7\% more sources overall) and greater astrometric precision. The set of sources in EDR3 is distinct from that of DR2 not only in that more sources are present; some may have disappeared or changed identifiers \citep{torra2021gaiaedr3sourcelist}. \textit{Gaia} Data Release 3 (DR3) will include the same sources as EDR3 but with additional data products.

\subsection{Cross-matching}

We use the same method employed in CW20, and here provide a brief review, pointing the reader to that work for a more detailed background. The aim is to automatically match CSPNe or compact/stellar-like nebulae to \textit{Gaia} sources. The expectation is that these sources will be close to the catalogued position of the PN, but the possibility of non-detections, background interlopers, and positional inaccuracies necessitates a more nuanced approach than simple nearest-neighbour matching.

The likelihood ratio method of \citet{sutherlandsaunders1992} is well-suited to this aim. Given a candidate match at angular separation $r$, the approach compares the probability of finding a genuine match at $r$ to finding a background object. Our implementation assumes a uniform background density in each field, and a radially symmetric positional distribution of genuine CSPNe parameterised by the radius of the PN (allowing larger PNe to have more offset CSPNe, albeit with lower probability). We also employ the colour distributions of true CSPNe and background sources, parameterised based on the photometric excess factor \citep{evans2018gaiaphotometry}, with the effect of down-weighting colours for sources whose colour is potentially contaminated by nearby sources or background. Distributions are derived iteratively, using sources with secure identifications based on colour to empirically determine the positional distribution of genuine CSPNe, and in turn using the positionally secure sources to update the colour distribution. This works because CSPNe are often much bluer than typical \textit{Gaia} sources and therefore can be selected purely based on colour with relatively high confidence.

We made a few small changes to the algorithm's treatment of the $G_\textup{BP}$--$G_\textup{RP}$ colours. In deriving the initial and final colour distributions, we only considered sources with $G$ brighter than 19, which have relatively precise colours. Then in computing the likelihood ratios for individual sources we convolved the densities with the photometric uncertainties, after accounting for the smoothing already applied by the kernel density estimation. This lessens the impact of faint sources with extreme and likely spurious colours. We also adopted the prescription of \citet{riello2021edr3photometry} for describing a locus of photometric excess values of well-behaved sources. The photometric excess is no longer used by \textit{Gaia} as a filter for published photometry, though for large excess values the photometry is still not very discriminative, as it likely dominated by the background.

We retrieved all \textit{Gaia} EDR3 sources within a 60\arcsec\ radius of the PN positions from HASH. All of these were used in calculating the background density and colour distribution (the latter only for $G$ < 19), while only the subset of sources within half each PN angular size (taken from HASH) plus 2\arcsec\ were considered as candidates. For each PN, the set of candidate likelihood ratios were used to calculate reliabilities (the probability that a given candidate is the correct match).

\section{Catalogue} \label{sec:results}

\begin{figure}
    \centering
    \includegraphics[width=0.49\hsize]{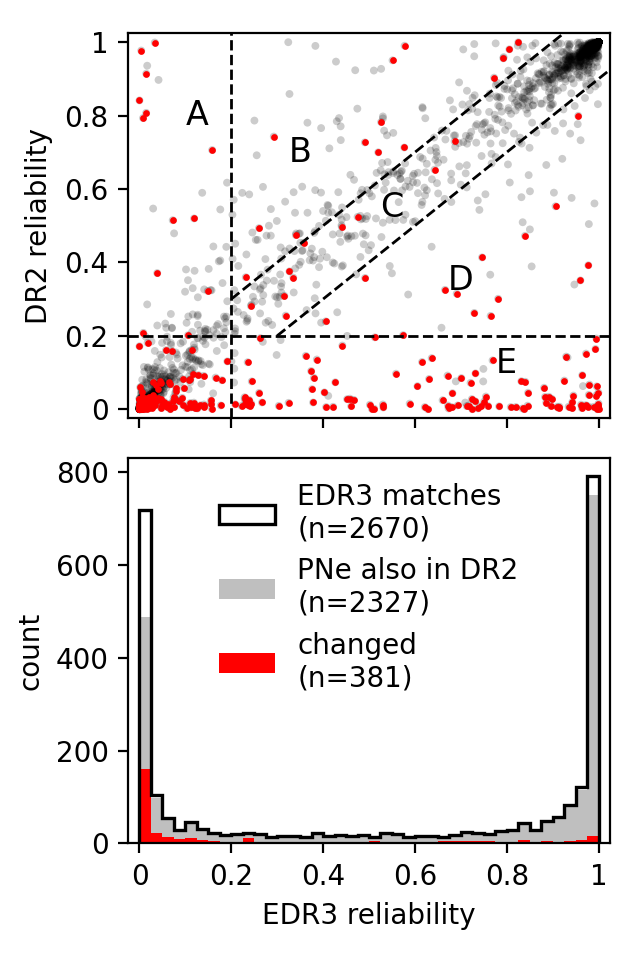}
    \includegraphics[width=0.49\hsize]{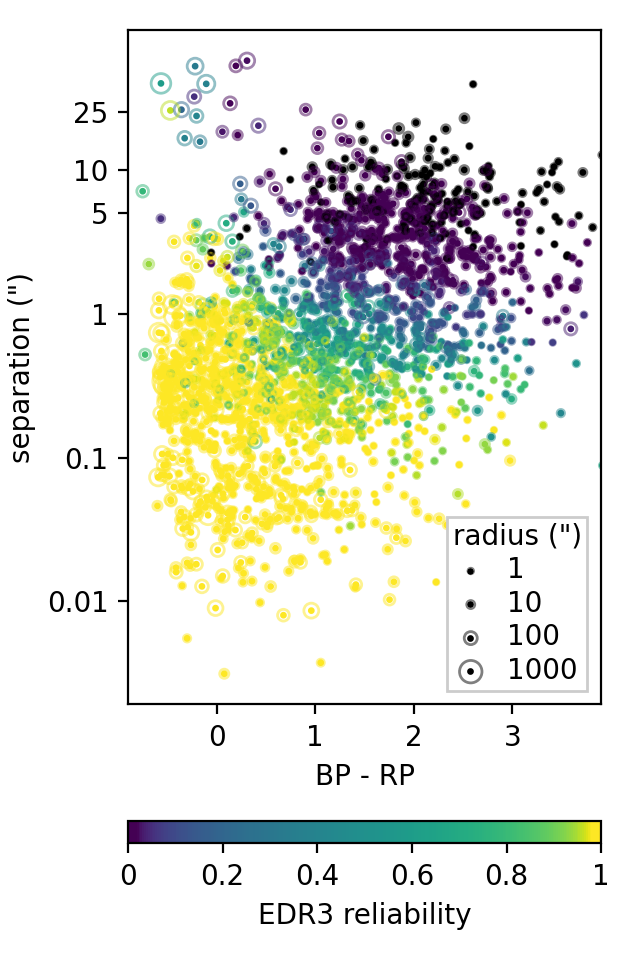}
    \caption{Matching results for "true" PNe. The top left panel shows the reliability of the CW20 \textit{Gaia} DR2 matches versus the EDR3 matches from this work for PNe present in both. Red points are PNe where the location of the best candidate match has changed by more than 0\farcs1. All of these points are contained in the histogram in the lower left panel, which also includes PNe that were not in CW20. The upper right panel shows the separation versus colour distribution of highest-ranked candidates, colour-coded by reliability with the non-linear scale shown at the bottom right. Rings around markers indicate PN angular sizes.}
    \label{fig:dr2_comparison}
\end{figure}

The improved method applied to EDR3 produces a bimodal distribution of top match reliabilities similar to that from CW20 for \textit{Gaia} DR2 (Fig \ref{fig:dr2_comparison}, lower left; for clarity only "true" PNe are shown). Most sources have reliabilities close to their values from CW20; these are near the diagonal (region C) in upper left of Fig. \ref{fig:dr2_comparison}. Points in regions A and E represent the most significant changes, typically due to removed or new sources respectively. Counts of PNe in these regions are in Table \ref{table:counts}, including "likely" and "possible" PNe. In addition there are 207 new PNe in HASH since CW20 (included in Fig. \ref{fig:dr2_comparison}; lower left); of these 107 have candidates with reliability > 0.2. There are also 22 objects published in CW20 that are no longer listed as PNe in HASH (not shown).

As with the previous iteration, the method retrieves blue CSPNe even out to large separations (Fig. \ref{fig:dr2_comparison}, right; the largest PN in the cluster at the upper left has an angular radius of nearly 600\arcsec). The few sources with unphysically blue colours are scored lower due to their large photometric uncertainties.

\begin{table}
\caption{\label{table:counts}Counts of PNe occupying different regions of the scatter plot in the upper left corner of Fig. \ref{fig:dr2_comparison}, representing different changes in reliability relative to the CW20 \textit{Gaia} DR2 matches.}
\centering
\begin{tabular}{lllllll}
\hline\hline
PN status & match location & \multicolumn{5}{c}{region}\\
& & A & B & C & D & E\\
\hline
true & same & 53 & 105 & 1238 & 47 & 15\\
 & changed & 16 & 14 & 10 & 14 & 157\\
other & same & 36 & 58 & 303 & 7 & 3\\
 & changed & 3 & 3 & 2 & 4 & 30\\
\hline
\end{tabular}
\end{table}

The complete catalogue of the 2117 highest ranked candidates with reliability > 0.2 is in Table A.1.
The catalogue includes the \textit{Gaia} source identifier corresponding to the best match for each PN, the reliability of that match, and its angular separation from the PN position from HASH. Copied into the catalogue are relevant data from HASH (name, PN G identifier, PN coordinates, angular size, and confirmation status) and \textit{Gaia} EDR3 (source coordinates, photometry, and astrometry). We additionally include the image parameter determination \citep[IPD,][]{lindegren2021edr3astrometry} goodness-of-fit harmonic amplitude from \textit{Gaia} (see Sect. \ref{sect:nebuladetect}) and combined statistical and parallax distance derived in this work (see Sect. \ref{sect:distancescombined}).

\subsection{Individual objects}

\begin{figure}
    \centering
    \begin{subfigure}[t]{\hsize}
        \centering
        \includegraphics[width=0.03\hsize]{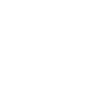}
        \includegraphics[width=0.23\hsize]{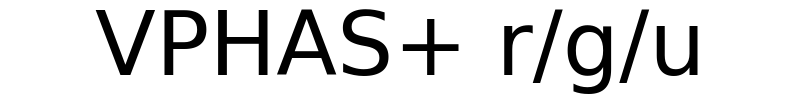}
        \includegraphics[width=0.23\hsize]{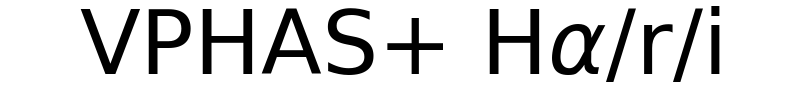}
        \includegraphics[width=0.23\hsize]{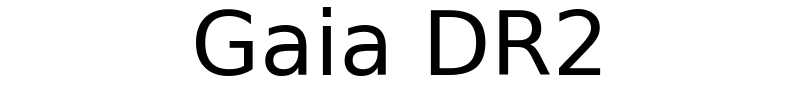}
        \includegraphics[width=0.23\hsize]{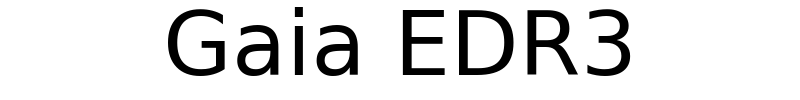}
    \end{subfigure}
    \par\vspace{-1.5pt}
    \begin{subfigure}[t]{\hsize}
        \centering
        \includegraphics[width=0.03\hsize]{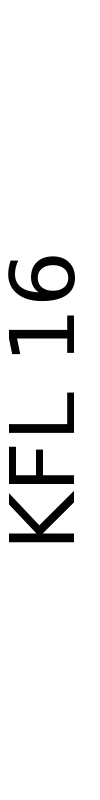}
        \includegraphics[width=0.23\hsize]{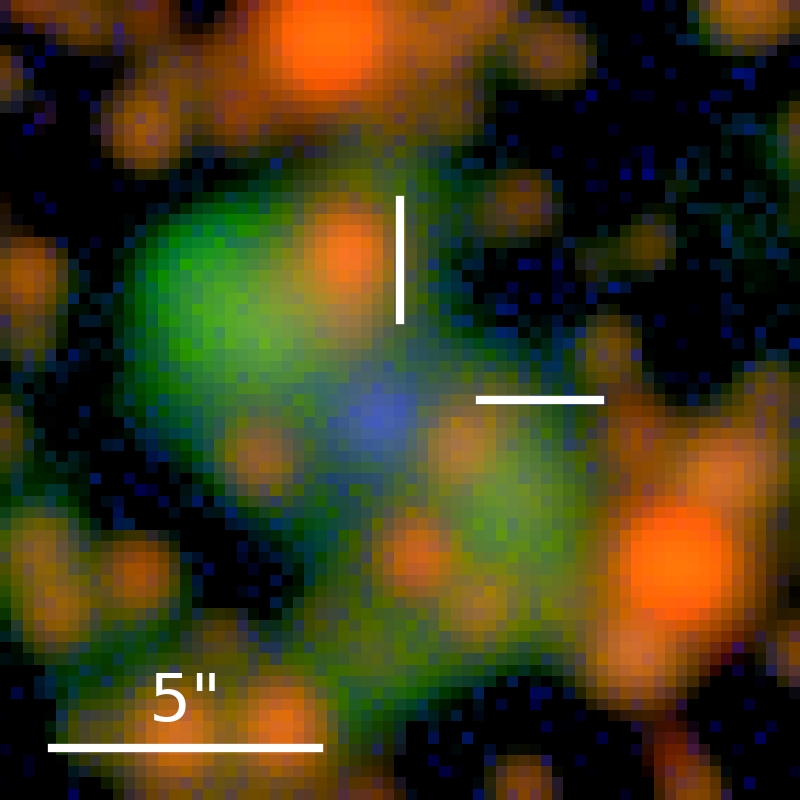}
        \includegraphics[width=0.23\hsize]{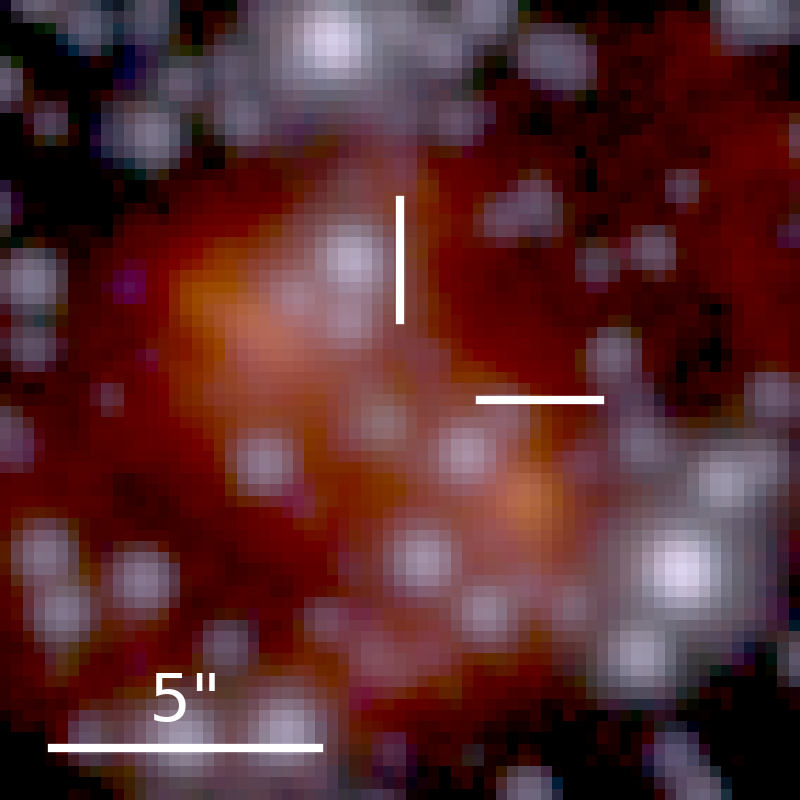}
        \includegraphics[width=0.23\hsize]{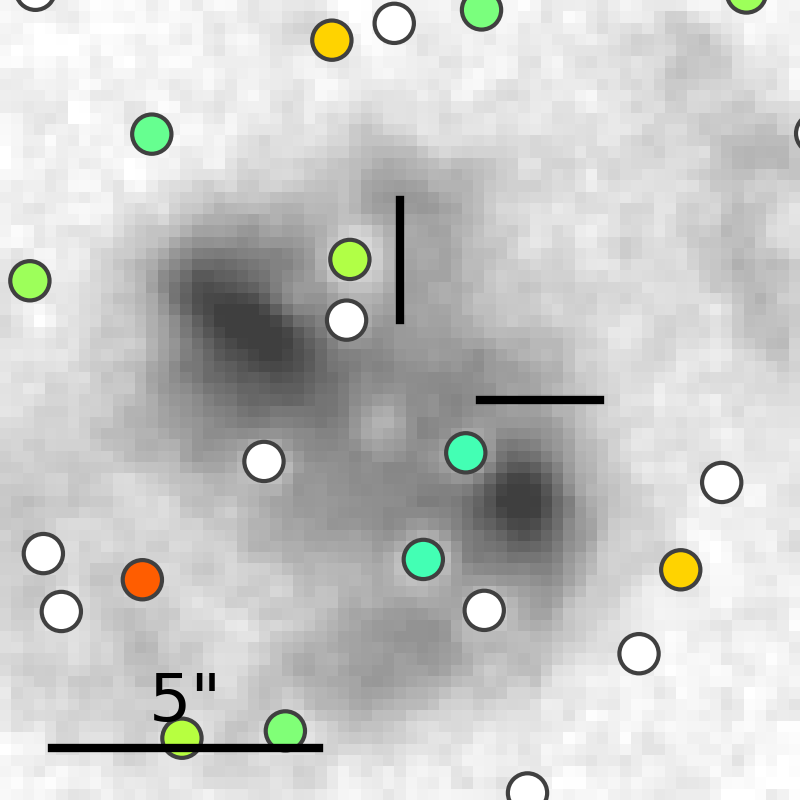}
        \includegraphics[width=0.23\hsize]{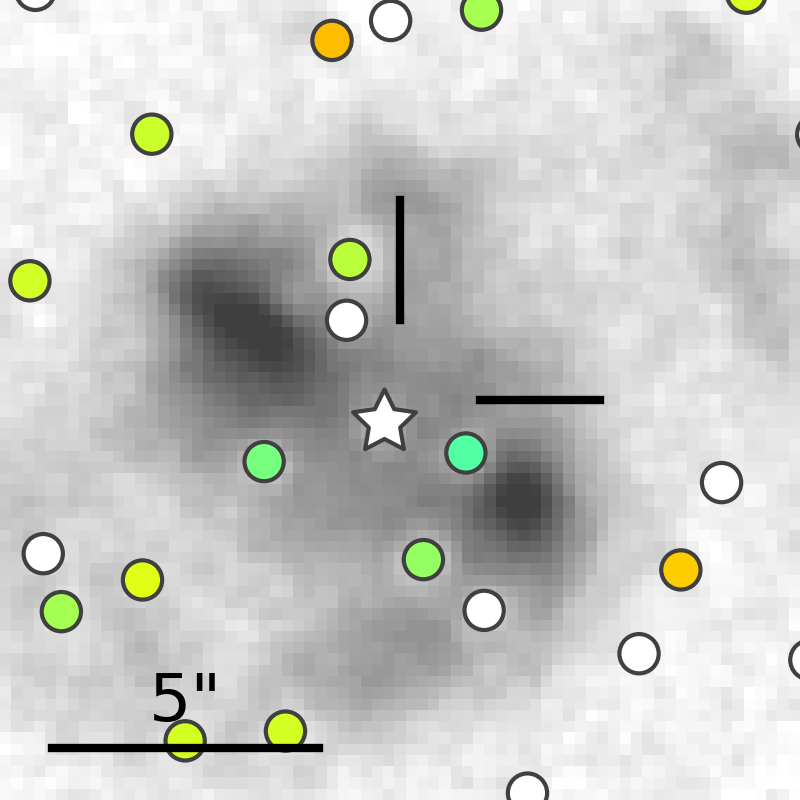}
    \end{subfigure}
    \par\vspace{-1.5pt}
    \begin{subfigure}[t]{\hsize}
        \centering
        \includegraphics[width=0.03\hsize]{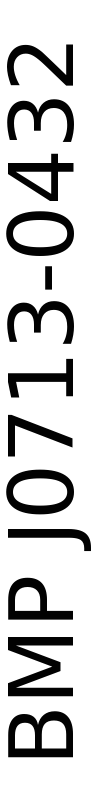}
        \includegraphics[width=0.23\hsize]{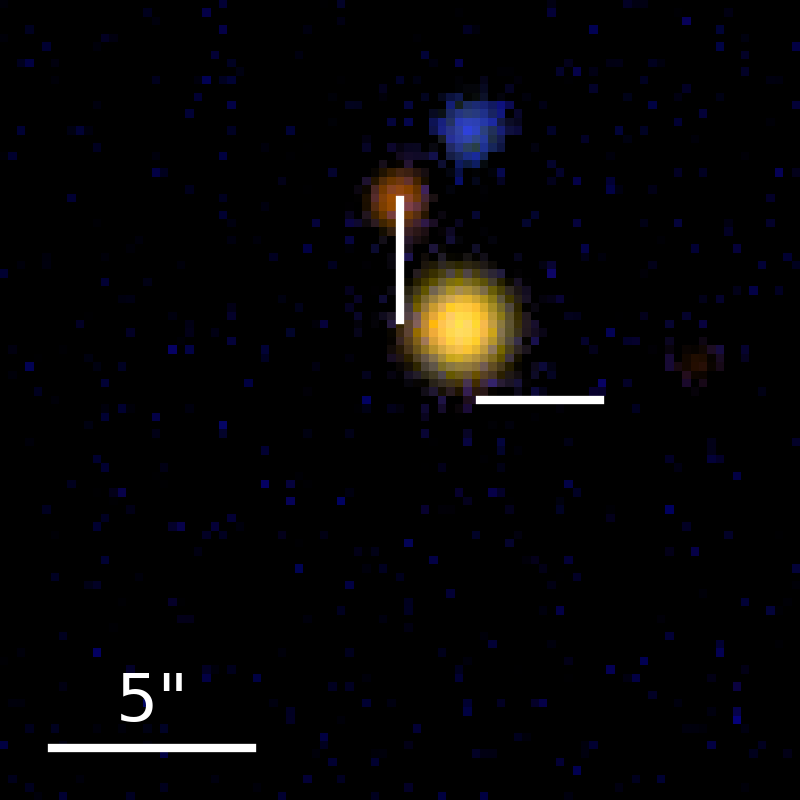}
        \includegraphics[width=0.23\hsize]{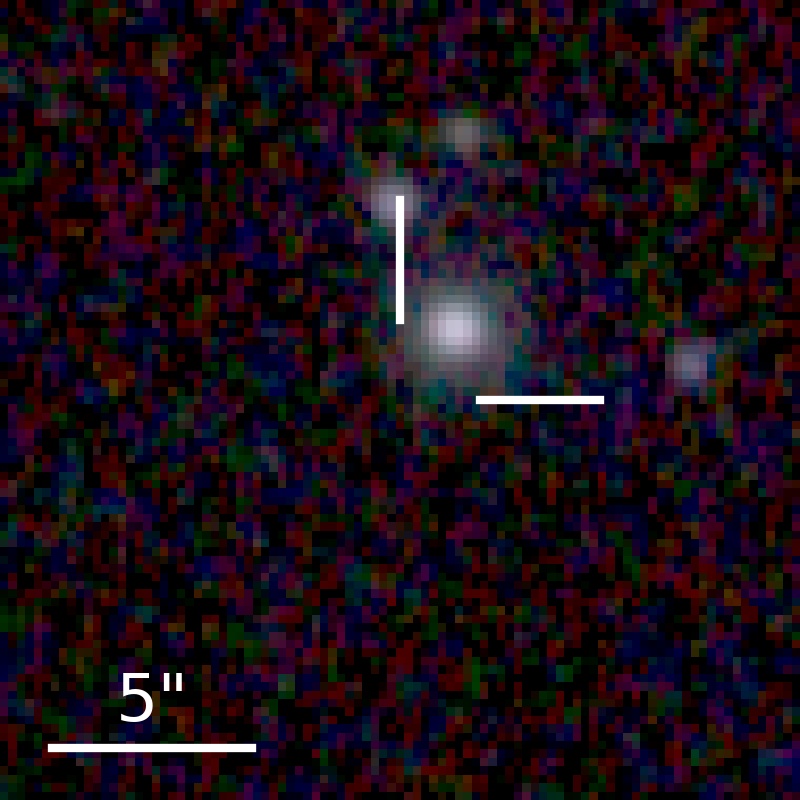}
        \includegraphics[width=0.23\hsize]{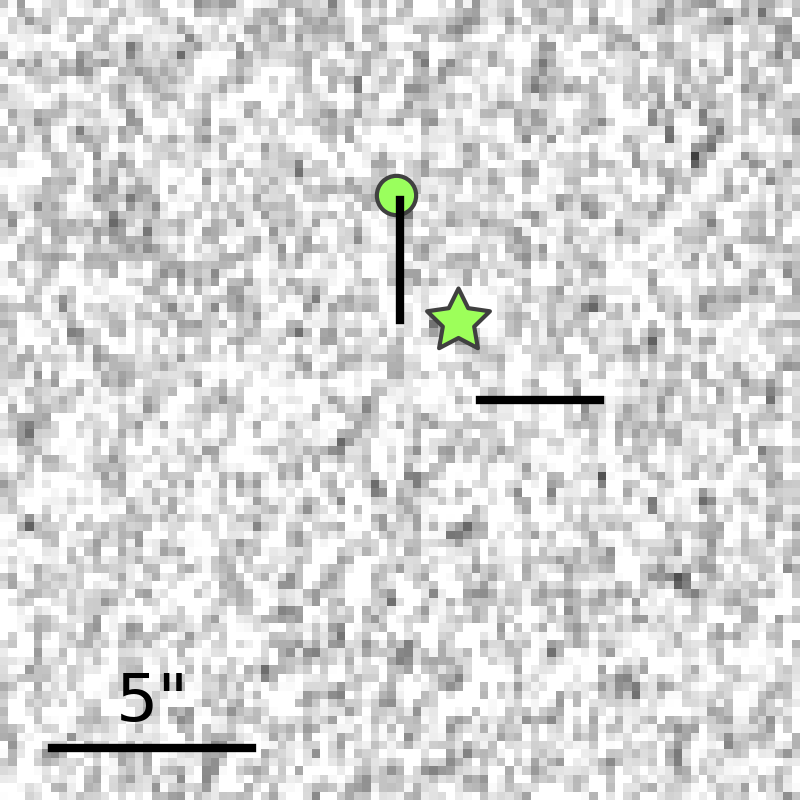}
        \includegraphics[width=0.23\hsize]{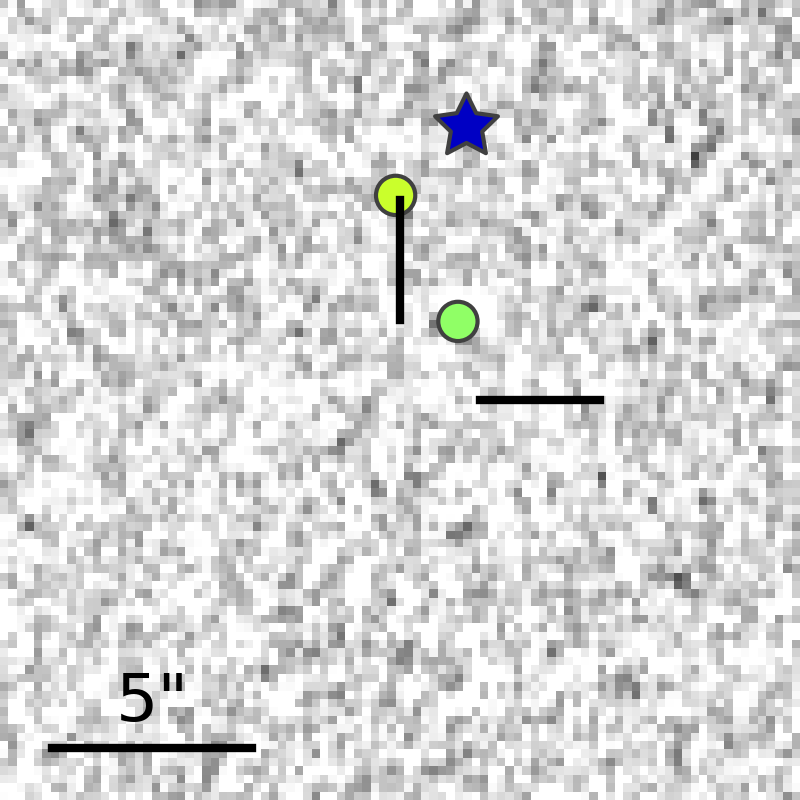}
    \end{subfigure}
    \par\vspace{-1.5pt}
    \begin{subfigure}[t]{\hsize}
        \centering
        \includegraphics[width=0.03\hsize]{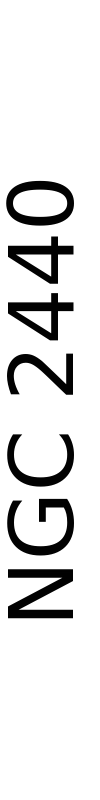}
        \includegraphics[width=0.23\hsize]{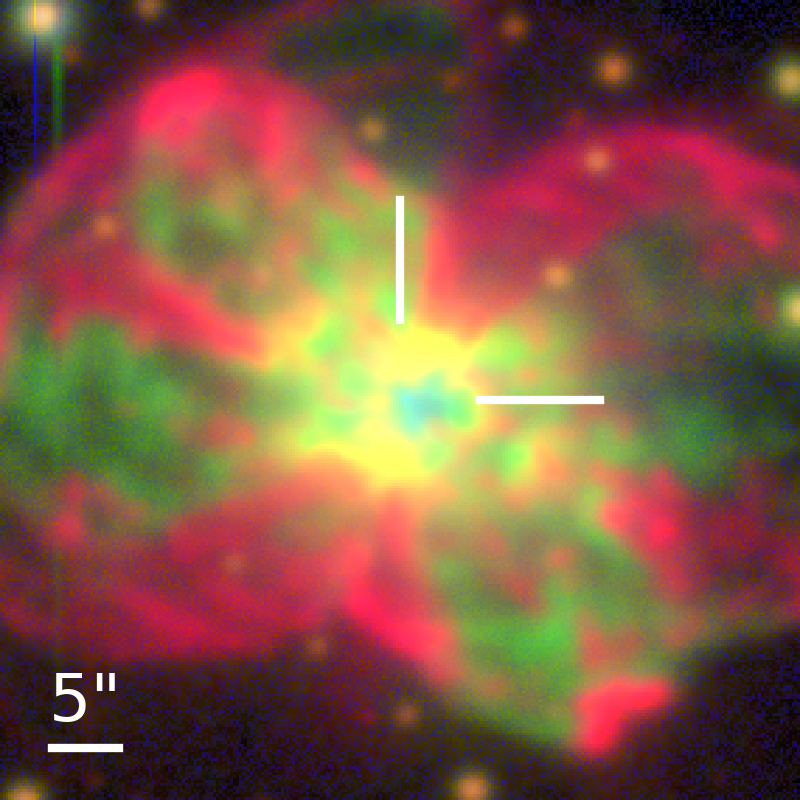}
        \includegraphics[width=0.23\hsize]{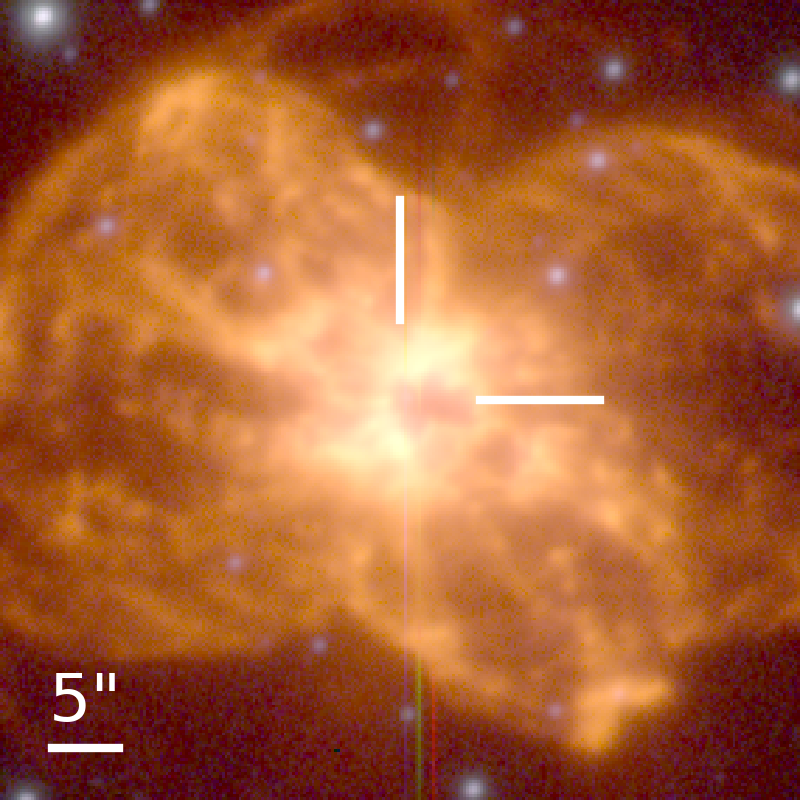}
        \includegraphics[width=0.23\hsize]{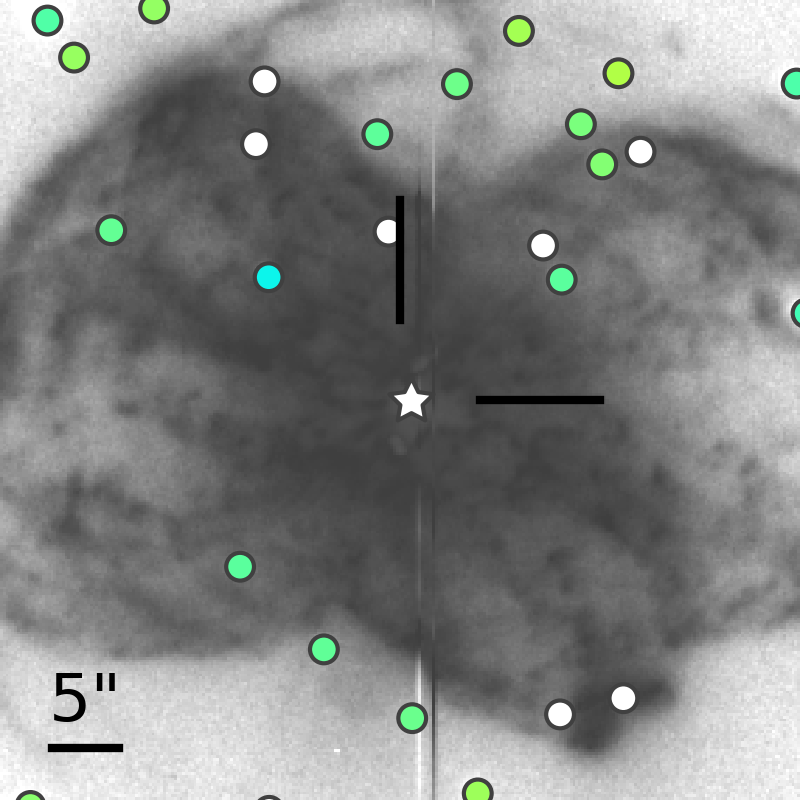}
        \includegraphics[width=0.23\hsize]{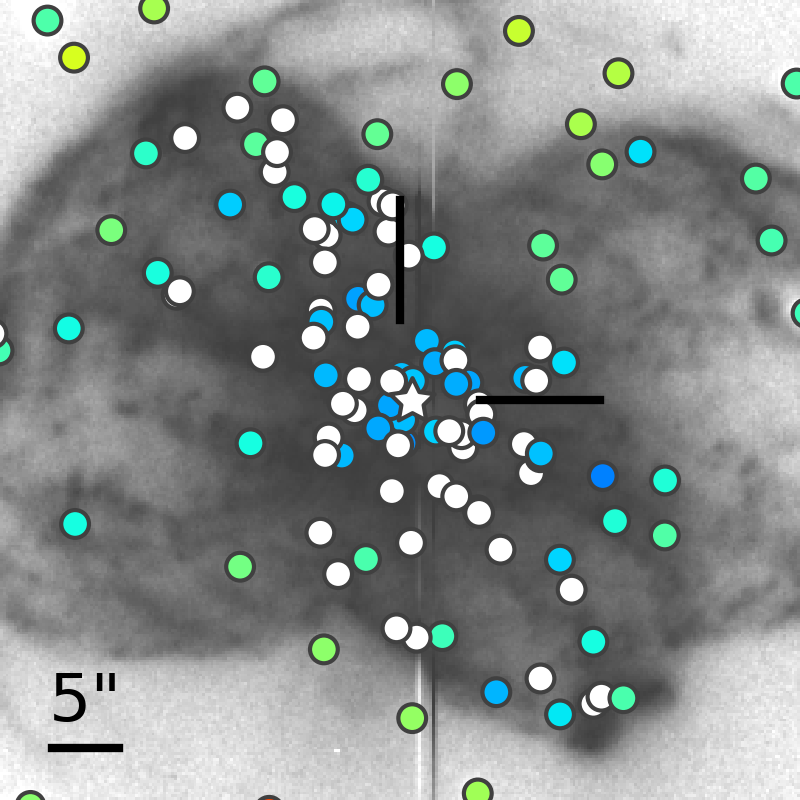}
    \end{subfigure}
    \par\vspace{-1.5pt}
    \begin{subfigure}[t]{\hsize}
        \centering
        \includegraphics[width=0.03\hsize]{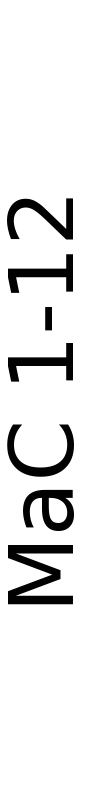}
        \includegraphics[width=0.23\hsize]{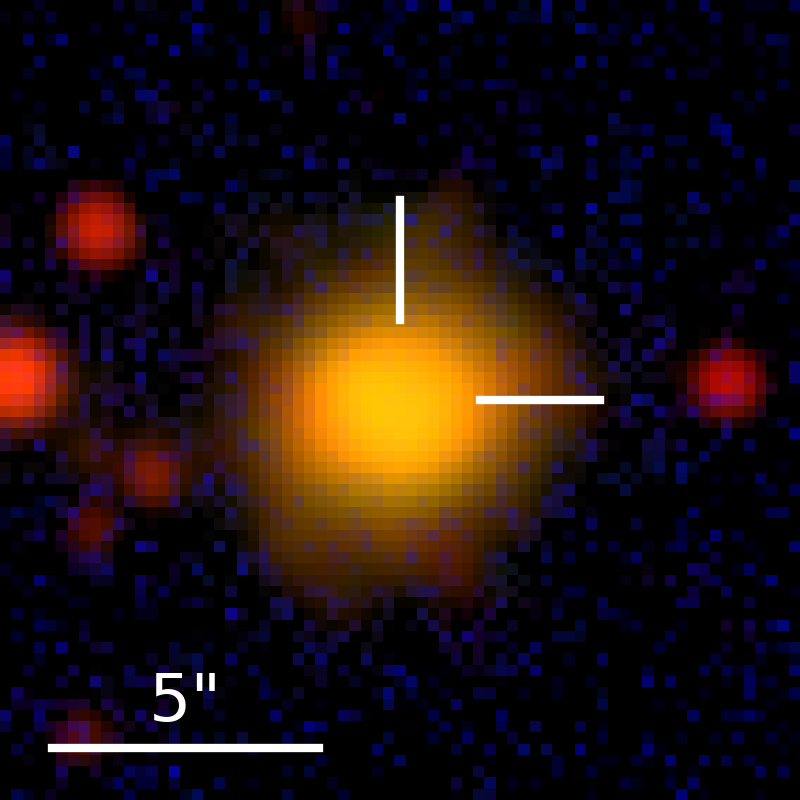}
        \includegraphics[width=0.23\hsize]{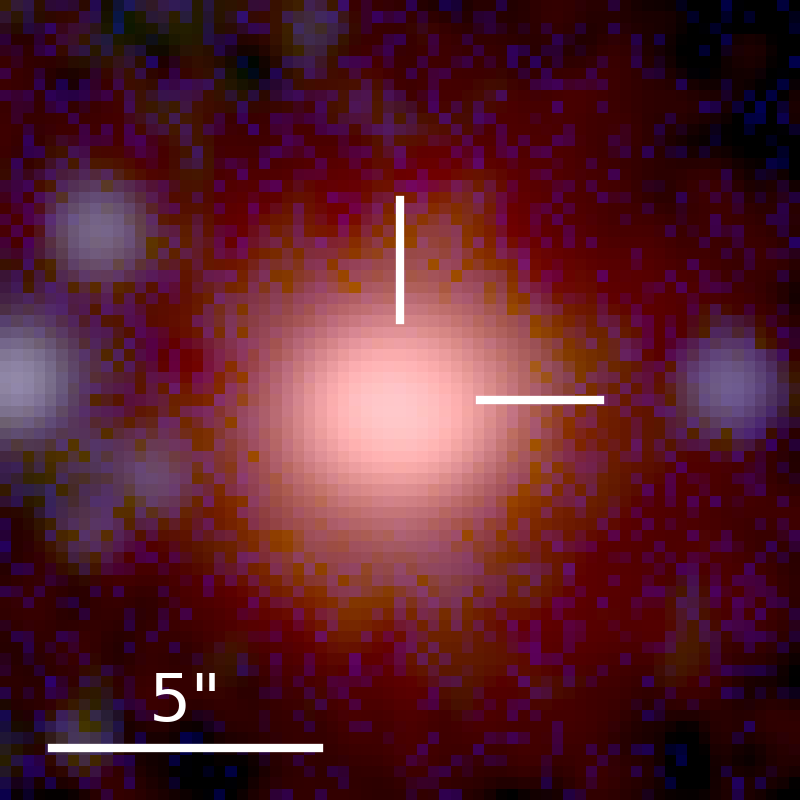}
        \includegraphics[width=0.23\hsize]{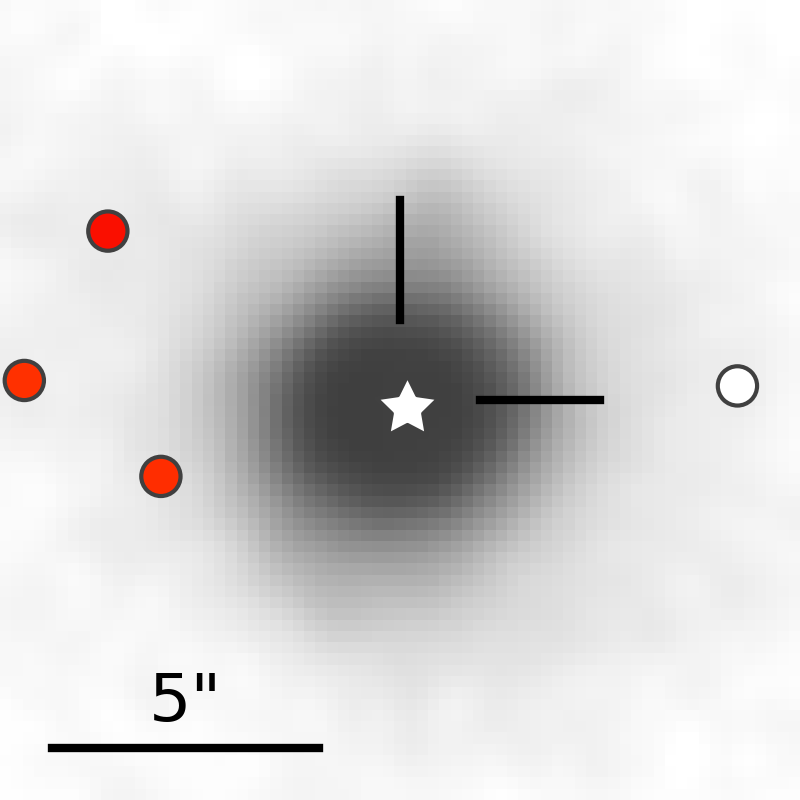}
        \includegraphics[width=0.23\hsize]{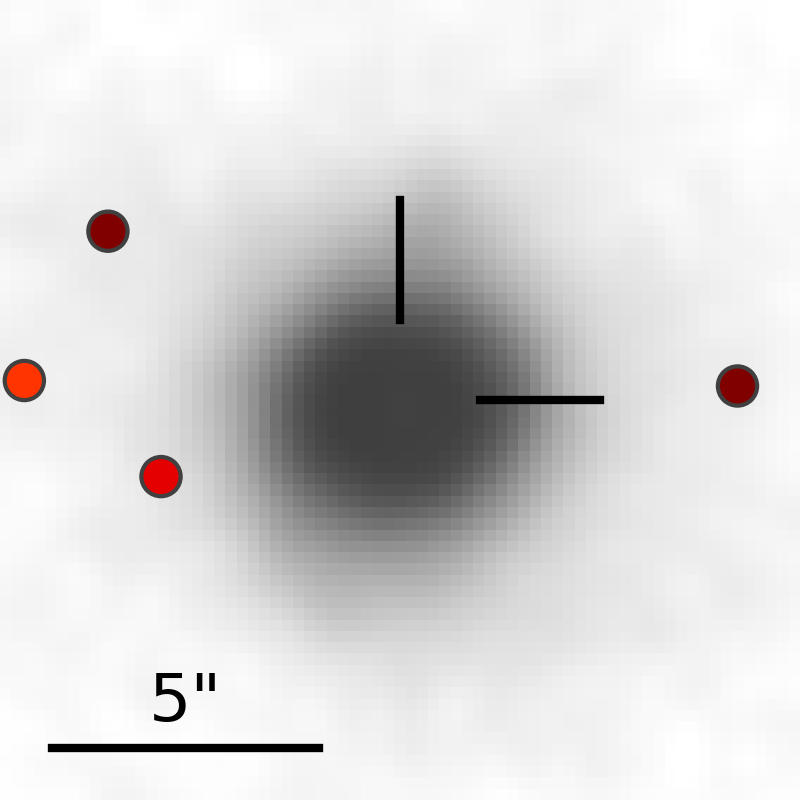}
    \end{subfigure}
    \par\vspace{-1.5pt}
    \includegraphics[width=0.03\hsize]{pne/textblank.png}
    \includegraphics[width=0.95\hsize]{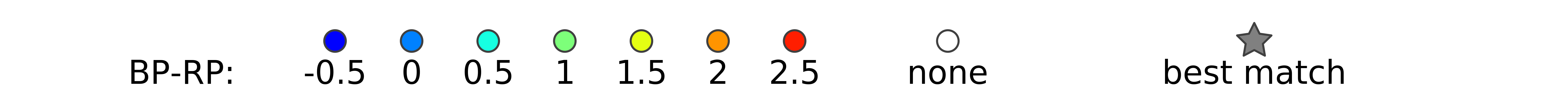}
    \caption{VPHAS+ images of select PNe centred on their coordinates from HASH. North is up and east is to the left.
    The left two columns are colour images in different sets of filters, while the right two columns are quotient ($r^\prime$ -- H$\alpha$) images overlayed with \textit{Gaia} sources coloured according to their $G_\textup{BP}$--$G_\textup{RP}$ colours, with the best match highlighted.}
    \label{fig:individualobjects}
\end{figure}

Fig. \ref{fig:individualobjects} shows wide- and narrow-band imagery from the VPHAS+ survey \citep{vphassurvey} overlayed with \textit{Gaia} EDR3 (and DR2) sources for four PNe chosen to exemplify significant changes encountered in \textit{Gaia} EDR3. Data from VPHAS+ were not used as part of the matching, but the images, which capture stellar colours and nebula emission, provide good examples of some of the changes encountered in the updated \textit{Gaia} data.

KFL 16 and BMP J0713-0432 both have new detections in EDR3. The CSPN of KFL 16 has no \textit{Gaia} colour but is sufficiently central that it is selected; its blue colour is evident in the VPHAS+ imagery. The CSPN of BMP J0713-0432 does exhibit a blue $G_\textup{BP}$--$G_\textup{RP}$ colour and as such is correctly identified despite not being the closest source to the PN position from HASH.

NGC 2440 had a clear CSPN detection in DR2. In EDR3 the central region of the PN now contains many nebula detections.
The true CSPN is still selected based on its position (it is overwhelmed by the bright nebula in VPHAS+ but clear in space-based imagery), but with reduced confidence on account of the many new nearby sources. In the case of MaC 1-12, the lone detection from DR2 has disappeared, and no matching EDR3 source is found.

\subsection{Visual binaries} \label{sect:visualbinaries}

Visual binaries are astrophysically interesting because, among other reasons, characterisation of the companion can provide an estimate of the distance to the pair and thus to the PN. However close companions - whether genuine or merely a chance alignment - can provide a challenge for our matching. Of the 53 PNe with two potential matches (two candidates with reliability > 0.2), about half are pairs with separation less than 1\arcsec. These likely include some genuine binary systems, though their blended colour photometry and degraded astrometry makes it difficult to characterise the companion star and secure its relation through astrometry \citep[e.g.][]{gonzalez2020widebinaries}.

Some CSPNe are visual binaries in \textit{Hubble} Space Telescope (HST) imagery: ten are presented in \citet{ciardullo1999hstbinaries} ("probable associations") as well as additional pairs in \citet{benetti2003hstbinaries} and \citet{liebert2013hstbinaries}. Typical separations of known pairs are fractions of an arcsecond; these appear blended from the ground. With the improved resolution of \textit{Gaia} EDR3, we expect that more of these will be resolved: the source separation limit has decreased to 0\farcs18 from 0\farcs4 in DR2. however incompleteness starts to set in below 1\farcs5 and is quite severe below 0\farcs7 \citep{fabricius2021edr3cataloguevalidation}.

The detections for these PNe are listed in Table \ref{table:visual}. Even sources that not separated by \textit{Gaia} can show signs of binarity if they have a nonzero IPD multipeak fraction, which indicates that some detections show multiple peaks. These can indicate a close visual binary, though depending on the separation it may only be resolved in some scan directions.

\begin{table}
\setlength{\tabcolsep}{4.7pt}
\caption{\label{table:visual}\textit{Gaia} detections of known close visual binaries.}
\centering
\begin{tabular}{lllllll}
\hline\hline
PN & Sep.\tablefootmark{a} & $\Delta V$\tablefootmark{b} & Rel.\tablefootmark{c} & $G$ &  Mul.\tablefootmark{d} &  RUWE\\
  &  (\arcsec)  & (mag) &  & (mag) &    (\%)     &               \\
\hline
Abell 31 & 0.26 & >6.8 & 1.00 & 15.5 & 0 & 1.03 \\ 
Abell 33 & 1.80\tablefootmark{e} & 1.0 & 0.93 & 15.9 & 0 & 1.09 \\
 &  & & 0.07 & 16.7 & 0 & 0.98 \\
EGB 6 & 0.17 & $\ldots$ & 0.96 & 16.0 & 0 & 2.09 \\
K 1-14 & 0.36 & 2.4 & 1.00 & 16.1 & 70 & 2.41 \\
K 1-22 & 0.35\tablefootmark{e} & 0.3 & 0.83 & 16.7 & 88 & 3.47 \\
 &  & & 0.17 & 16.7 & 79 & 2.94 \\
K 1-27 & 0.56 & 5.1 & 1.00 & 16.0 & 0 & 0.90 \\
Mz 2 & 0.28 & -1.3 & 0.98 & 16.8 & 0 & 0.97 \\
NGC 1535 & 1.04 & 5.5 & 1.00 & 12.1 & 17 & 1.01 \\
NGC 3132 & 1.70\tablefootmark{e} & -5.7\tablefootmark{f} & 0.81 & 10.0 & 0 & 1.84 \\ 
 & & & 0.19 & 16.1 & 16 & 2.07 \\
NGC 6818 & 0.09 & 0.7 & 1.00 & 16.3 & 0 & 3.80 \\ 
NGC 7008 & 0.42 & -1.5 & 1.00 & 13.7 & 93 & 6.07 \\ 
Sp 3 & 0.31 & -3.7 & 1.00 & 13.1 & 45 & 3.02 \\
\hline
\end{tabular}
\tablefoot{Rows with first columns blank indicate the
\textit{Gaia}-detected companion of the (assumed PN progenitor) in the row above. Single rows indicate that \textit{Gaia} has not resolved the pair.
\tablefoottext{a}{Literature separation, except where noted. Most values are from \citet{ciardullo1999hstbinaries}, except for EGB 6 \citep{liebert2013hstbinaries} and NGC 6818 \citep{benetti2003hstbinaries}.}
\tablefoottext{b}{Difference in HST $V$ magnitude between ionising star and companion (a positive difference indicates companion is fainter).}
\tablefoottext{c}{Reliability.}
\tablefoottext{d}{IPD multipeak fraction.}
\tablefoottext{e}{\textit{Gaia} value, within 0\farcs02 of literature HST value.}
\tablefoottext{f}{The ionising star of NGC 3132 is thought to be the fainter of the pair, so highest reliability match here is not correct in that sense.}
}
\end{table}

Abell 33 and NGC 3132 have companions at wide separations detectable from the ground (though the catalogue position of NGC 3132 is actually closer to that of the companion star, leading to ambiguity in our matching results). Of the closer pairs, only K 1-22 is resolved into separate sources by \textit{Gaia}, possibly on account of their similar magnitudes. Another four PNe have a notable fraction of multi-peak detections, while the remaining PNe seem to have companions that are either too faint (e.g. Abell 31 and K 1-27) or too close (e.g. NGC 6818). The re-normalised unit weight error \citep[RUWE, indicative of excess astrometric error;][]{gaiaruwe} values range from normal to significant, and none are labelled as duplicated sources.

\subsection{Nebula detections} \label{sect:nebuladetect}

\begin{figure}
    \centering
    \includegraphics[width=0.45\hsize]{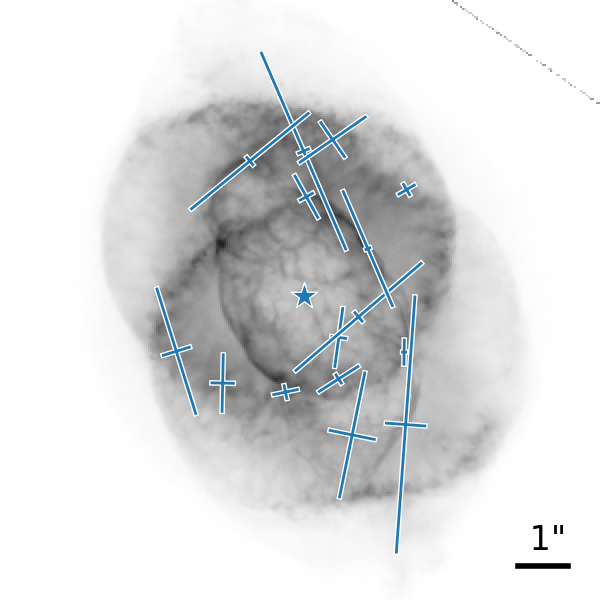}
    \includegraphics[width=0.45\hsize]{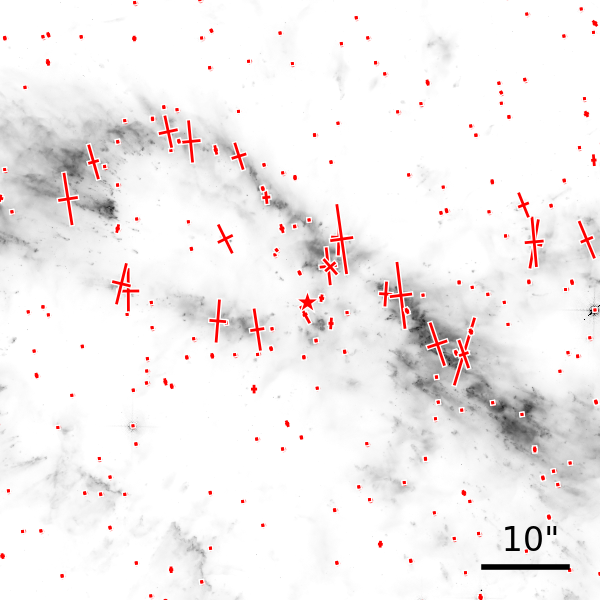}
    \includegraphics[width=0.98\hsize]{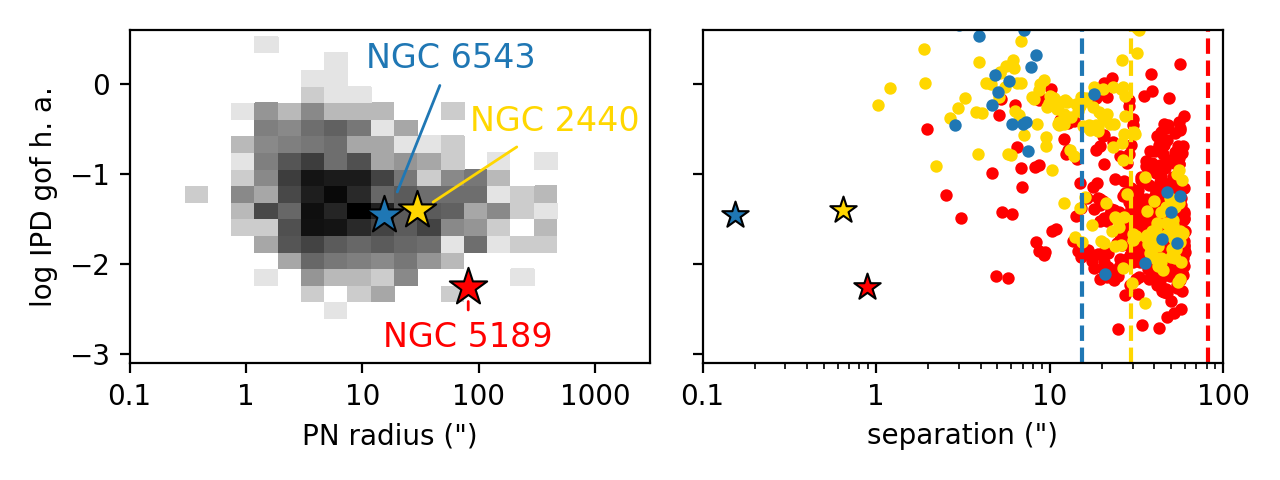}
    \caption{Nebula detections for bright PNe. The upper images are HST narrowband images of NGC 6543 (left) and the central region of NGC 5189 (right) overlayed with \textit{Gaia} EDR3 sources (coloured points and crosses). Error bars have been rotated to reflect the direction of the principle component of the error covariance matrices, and scaled by a factor of 100 for visibility.
    The lower plots show the distribution of IPD goodness-of-fit harmonic amplitudes. The left plot shows the distribution for the best matches versus PN angular radius on a linear colourmap for all best matches in our catalogue, with locations of the CSPNe of the PNe above (and NGC 2440 from Fig. \ref{fig:individualobjects}) highlighted. The right plot shows all sources in the vicinity of these PNe, plotted against angular separation on a log scale. Vertical lines indicate the PN radii; the lack of points past 60\arcsec\ is a result of our selection cutoff.
    (Images credit: \textit{Hubble} Legacy Archive/ESA/NASA.)}
    \label{fig:nebula_detections}
\end{figure}

It is expected that for compact PNe \textit{Gaia} may not detect the CSPN but rather the bright central region of the nebula. \textit{Gaia} can also detect features in extended PNe; this was seen early on in \textit{Gaia}'s mission for
NGC 6543 \citep{fabricius2016gaiadr1sourcelist}.\footnote{Also featured as \textit{Gaia}'s image of the week: \url{https://www.cosmos.esa.int/web/gaia/iow_20141205}.} These detections were filtered out in the first two data releases, leaving only the 11th magnitude CSPN.
Now in EDR3 some
have returned \citep[Fig. \ref{fig:nebula_detections}, upper left; c.f. Fig. 14 in][]{fabricius2016gaiadr1sourcelist}.

A strong indicator that these sources are non-stellar is the large values of IPD goodness-of-fit harmonic amplitude (which we abbreviate to HA). This
is newly included in \textit{Gaia} EDR3, and measures how the astrometric goodness-of-fit varies with scan angle. A large HA suggests an elongated source, such as a galaxy or a partially resolved binary.\footnote{HA values for the close visual binaries in the previous section are smaller than those of typical nebula detections;
the largest is 0.26 for NGC 6818, the closest pair in the sample.}

The lower left panel of Fig. \ref{fig:nebula_detections} shows the HA distribution for all of our high-confidence matches ("true" PN with reliability > 0.8), parameterised by the angular size of the PN. Many compact PNe have larger HA values indicating likely nebular detections (compared to the values for extended PNe, whose detections we expect to be stellar). The lower right panel shows the distribution of all candidate matches for a few select objects. There is a clear separation between the stellar CSPN, nearby nebula detections (sources within the radius of the PN having high HA), and (usually more distant) field stars.

Nebula detections of extended PN appear rare, and to keep our approach simple, we do not use the HA in our matching. We do however include it in our catalogue as an approximate way to check how stellar-like a source is. Large values should be treated with caution, particularly in conjunction with a high photometric excess factor, but do not necessarily indicate a nebula detection.

We note that so far these detections are not useful for studies of nebular evolution. NGC 6543 for example has a measured expansion rate of $3.13 \pm 0.16$ mas yr$^{-1}$ \citep[similar to other PNe;][]{expansion2018}, large enough for \textit{Gaia} to detect as proper motion. However the nebula detections have only two-parameter solutions (positions only) with uncertainties much larger than any change due to expansion over \textit{Gaia}'s mission duration.

\section{Distances} \label{sect:distances}

Distances to Galactic PNe have long been a key challenge in the study of PNe, with most distances reliant on statistical relations such as the H$\alpha$ surface brightness to physical radius relation of \citet[][hereinafter FPB16]{frewsurfacebrightness2016}. The precision of the parallaxes in \textit{Gaia} EDR3 should translate into improved precision in distance determinations, however leveraging these parallaxes is not entirely straightforward.

\subsection{Parallaxes compared to statistical distances}

Following \citet{smith2015}, parallaxes can be used to evaluate the accuracy of a statistical distance scale using distance ratios, that is, the per-object product of statistical distance and parallax. An accurate statistical relation should yield a distance ratio distribution centred on unity. We use this method to compare \textit{Gaia} parallaxes to the statistical distance scale of FPB16. Parallaxes are corrected for the zero point bias following \citet{lindegren2021parallaxzeropoint}. We use the sub-trend relations where applicable, and the same quality cuts as in CW20, with the aim of selecting high quality measurements without biasing the sample ("true" PNe, reliability > 0.98, \texttt{visibility\_periods\_used} > 8, RUWE < 1.4, parallax error < 0.2 mas).

We find a median ratio of $1.03\pm0.02$ and a mean ratio of $1.11\pm0.04$ (uncertainties calculated via bootstrap). The overall distance scale is matched well (with some suggestion of the distances being slightly high), now with nearly twice as many objects being considered (294 compared to 160 in CW20). There is no statistically significant trend versus the PN surface brightness, implying the slope of the relation is reasonably accurate.

The effect of the parallax zero point correction (typically on the order of the median quasar parallax, -0.017 mas) is to make most parallaxes larger (implying smaller distances). Without the correction the distance ratios would have a smaller median value of 0.97, and exhibit a small but significant trend with surface brightness, even after accounting for outliers. This demonstrates the need for caution interpreting parallaxes, particularly as a population in which systematics can have noticeable effects.

\subsection{Statistical distances as a prior} \label{sect:distancescombined}

Estimating distances from parallaxes is well known to require a proper prior \citep{xlurigaiaparallaxes}. \citet[][hereinafter BJ21]{bailerjones2021distances} adopted two different priors for their catalogue covering most of \textit{Gaia} EDR3: one based on a galaxy model and another that additionally incorporates \textit{Gaia}'s photometry.
Naturally, the effect of the adopted prior is greatest for stars with relatively large parallax uncertainties, which tends to result in distant, faint stars having their distances underestimated. The photometry prior aims to mitigate such effects. The rapid evolution of CSPNe makes such photometric distances less useful. However the relation between H$\alpha$ surface brightness and physical size of the PN used in FPB16 can perform a similar function.

For each PN, we use the statistical distance of FPB16 and its uncertainty as a prior $P(d)$, having shown in the previous section that the FPB16 scale remains broadly consistent with the \textit{Gaia} parallaxes. The \textit{Gaia} parallax $\omega$ and its uncertainty $\sigma_\omega$ provide the likelihood $P(\omega|d,\sigma_\omega)$. We compute the posterior distance distribution on a fixed grid of 1pc steps between 0 and 60kpc (twice the largest prior distance), by simply applying Bayes' theorem, that is, taking at each distance $d$ the posterior probability to be the product $P(d)P(\omega|d,\sigma_\omega)$, normalised by the sum of these values over the domain of the prior.

Following the catalogue format adopted by BJ21, we summarise the posterior by its median and 16th and 84th percentiles. We publish these values in our catalogue for all 733 sources with both parallaxes and statistical distances, regardless of reliability and excess astrometric noise. Both of those should be considered when adopting the distances. There are an additional 131 sources in our catalogue that lack statistical distances but have relative parallax errors better than 20\%; we do not publish distances for these but expect the distances from BJ21 to not be overly influenced by their choice of prior.

\begin{figure}
    \centering
    \includegraphics[width=0.98\hsize]{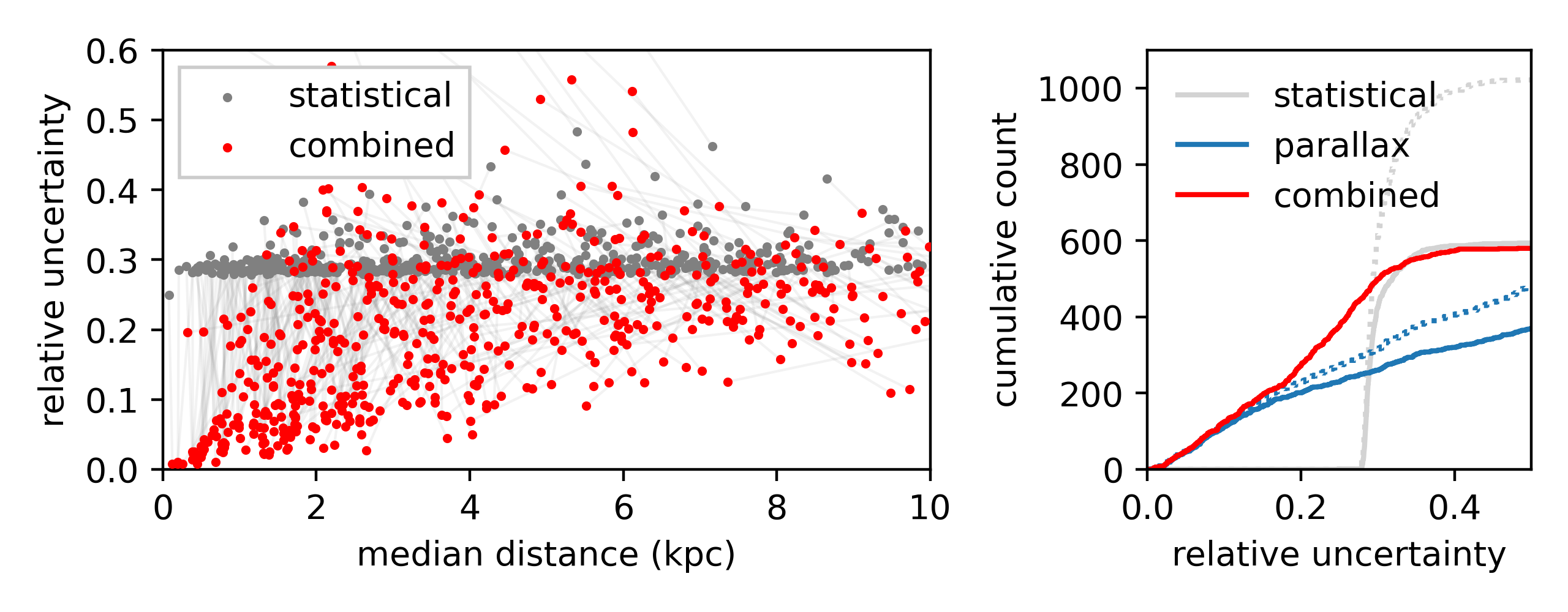}
    \caption{Improvement over statistical distances of combined distances for "true" PNe with high-reliability matches. The left panel shows the distribution of statistical and combined distances and relative uncertainties, with lines connecting pairs of points corresponding to the same object. The right panel shows the cumulative distribution of relative uncertainties for parallaxes, statistical distances, and combined distances. The dotted lines for parallaxes and statistical distances are counts including objects with only parallaxes (not in FPB16) and all "true" PNe from FPB16 (not necessarily having parallaxes or \textit{Gaia} matches) respectively.}
    \label{fig:distances}
\end{figure}

Treating the width of the distributions as the difference between the 16th and 84th percentiles, the median precision improvement (relative to the median distance) over the statistical distance prior is 1.4, with one third of the sample having its relative uncertainties improved by a factor of two or more (Fig. \ref{fig:distances}, left). A small fraction of sources have less concentrated posteriors, typically in the case where the prior and posterior modes have a large separation. The parallaxes dominate the uncertainties below relative errors of about 0.15 (Fig. \ref{fig:distances}, right).

The advantage of our adopted distances is that they leverage the \textit{Gaia} parallaxes while smoothly transitioning to statistical distances in cases where parallaxes are unavailable or extremely uncertain. For simplicity we have only considered the mean relation of FPB16; a similar approach could be taken incorporating the sub-trends in FPB16 or even the Galactic model of BJ21. The most appropriate prior will of course depend on the application.

\section{Conclusions} \label{sect:conclusions}

In this work we have presented a new version of our catalogue of candidate CSPNe and compact nebula detections, updated for \textit{Gaia} EDR3 and including newly identified PNe from HASH. While we have made small improvements to the matching algorithm, the minimal changes required demonstrate the robustness of our original method. We have also discussed the effects of close companions and nebula detections, and introduced a novel approach to combining PN statistical distances and \textit{Gaia} parallaxes.

The next \textit{Gaia} data release, \textit{Gaia} DR3, is expected in 2022. It is based on the exact same astrometric and photometric data as \textit{Gaia} EDR3. Hence the astrometry from this \textit{Gaia} EDR3 release will remain the best available for studying PNe based on their distances and kinematics until the full nominal mission \textit{Gaia} data release in several years to come.\footnote{See  \url{https://www.cosmos.esa.int/web/gaia/release} for the official release schedule.}

\begin{acknowledgements}
{
We would like to thank the referee, W. A. Weidmann, for his comments, which have helped improve the contents and clarity of this paper.

This research has made use of data from the European Space Agency (ESA) mission \textit{Gaia} (\url{https://www.cosmos.esa.int/gaia}), processed by the {\it Gaia} Data Processing and Analysis Consortium (DPAC; \url{https://www.cosmos.esa.int/web/gaia/dpac/consortium}). Funding for the DPAC has been provided by national institutions, in particular the institutions participating in the {\it Gaia} Multilateral Agreement.

This research has also made use of the HASH PN database (\url{http://hashpn.space}), and of Astropy (\url{http://www.astropy.org}), a community-developed core Python package for Astronomy \citep{astropy:2013, astropy:2018}.

Parts of this research were based on data products from observations made with ESO Telescopes at the La Silla Paranal Observatory under programme ID 177.D-3023, as part of the VST Photometric H$\alpha$ Survey of the southern Galactic plane and bulge (VPHAS+, \url{www.vphas.eu}).

This research was supported through the Cancer Research UK grant A24042.}
\end{acknowledgements}

\bibliographystyle{aa} 
\bibliography{bib.bib} 

\begin{thebibliography}{28}
\expandafter\ifx\csname natexlab\endcsname\relax\def\natexlab#1{#1}\fi

\bibitem[{{Astropy Collaboration} {et~al.}(2013){Astropy Collaboration},
  {Robitaille}, {Tollerud}, {Greenfield}, {Droettboom}, {Bray}, {Aldcroft},
  {Davis}, {Ginsburg}, {Price-Whelan}, {Kerzendorf}, {Conley}, {Crighton},
  {Barbary}, {Muna}, {Ferguson}, {Grollier}, {Parikh}, {Nair}, {Unther},
  {Deil}, {Woillez}, {Conseil}, {Kramer}, {Turner}, {Singer}, {Fox}, {Weaver},
  {Zabalza}, {Edwards}, {Azalee Bostroem}, {Burke}, {Casey}, {Crawford},
  {Dencheva}, {Ely}, {Jenness}, {Labrie}, {Lim}, {Pierfederici}, {Pontzen},
  {Ptak}, {Refsdal}, {Servillat}, \& {Streicher}}]{astropy:2013}
{Astropy Collaboration}, {Robitaille}, T.~P., {Tollerud}, E.~J., {et~al.} 2013,
  \aap, 558, A33

\bibitem[{{Bailer-Jones} {et~al.}(2021){Bailer-Jones}, {Rybizki}, {Fouesneau},
  {Demleitner}, \& {Andrae}}]{bailerjones2021distances}
{Bailer-Jones}, C.~A.~L., {Rybizki}, J., {Fouesneau}, M., {Demleitner}, M., \&
  {Andrae}, R. 2021, \aj, 161, 147

\bibitem[{{Benetti} {et~al.}(2003){Benetti}, {Cappellaro}, {Ragazzoni},
  {Sabbadin}, \& {Turatto}}]{benetti2003hstbinaries}
{Benetti}, S., {Cappellaro}, E., {Ragazzoni}, R., {Sabbadin}, F., \& {Turatto},
  M. 2003, \aap, 400, 161

\bibitem[{{Chornay} \& {Walton}(2020)}]{chornay2020cspn}
{Chornay}, N. \& {Walton}, N.~A. 2020, \aap, 638, A103

\bibitem[{{Ciardullo} {et~al.}(1999){Ciardullo}, {Bond}, {Sipior}, {Fullton},
  {Zhang}, \& {Schaefer}}]{ciardullo1999hstbinaries}
{Ciardullo}, R., {Bond}, H.~E., {Sipior}, M.~S., {et~al.} 1999, \aj, 118, 488

\bibitem[{{Drew} {et~al.}(2014){Drew}, {Gonzalez-Solares}, {Greimel}, {Irwin},
  {K{\"u}pc{\"u} Yoldas}, {Lewis}, {Barentsen}, {Eisl{\"o}ffel}, {Farnhill},
  {Martin}, {Walsh}, {Walton}, {Mohr-Smith}, {Raddi}, {Sale}, {Wright},
  {Groot}, {Barlow}, {Corradi}, {Drake}, {Fabregat}, {Frew}, {G{\"a}nsicke},
  {Knigge}, {Mampaso}, {Morris}, {Naylor}, {Parker}, {Phillipps}, {Ruhland},
  {Steeghs}, {Unruh}, {Vink}, {Wesson}, \& {Zijlstra}}]{vphassurvey}
{Drew}, J.~E., {Gonzalez-Solares}, E., {Greimel}, R., {et~al.} 2014, \mnras,
  440, 2036

\bibitem[{{Evans} {et~al.}(2018){Evans}, {Riello}, {De Angeli}, {Carrasco},
  {Montegriffo}, {Fabricius}, {Jordi}, {Palaversa}, {Diener}, {Busso},
  {Cacciari}, {van Leeuwen}, {Burgess}, {Davidson}, {Harrison}, {Hodgkin},
  {Pancino}, {Richards}, {Altavilla}, {Balaguer-N{\'u}{\~n}ez}, {Barstow},
  {Bellazzini}, {Brown}, {Castellani}, {Cocozza}, {De Luise}, {Delgado},
  {Ducourant}, {Galleti}, {Gilmore}, {Giuffrida}, {Holl}, {Kewley}, {Koposov},
  {Marinoni}, {Marrese}, {Osborne}, {Piersimoni}, {Portell}, {Pulone},
  {Ragaini}, {Sanna}, {Terrett}, {Walton}, {Wevers}, \&
  {Wyrzykowski}}]{evans2018gaiaphotometry}
{Evans}, D.~W., {Riello}, M., {De Angeli}, F., {et~al.} 2018, \aap, 616, A4

\bibitem[{{Fabricius} {et~al.}(2016){Fabricius}, {Bastian}, {Portell},
  {Casta{\~n}eda}, {Davidson}, {Hambly}, {Clotet}, {Biermann}, {Mora},
  {Busonero}, {Riva}, {Brown}, {Smart}, {Lammers}, {Torra}, {Drimmel},
  {Gracia}, {L{\"o}ffler}, {Spagna}, {Lindegren}, {Klioner}, {Andrei}, {Bach},
  {Bramante}, {Br{\"u}semeister}, {Busso}, {Carrasco}, {Gai}, {Garralda},
  {Gonz{\'a}lez-Vidal}, {Guerra}, {Hauser}, {Jordan}, {Jordi}, {Lenhardt},
  {Mignard}, {Messineo}, {Mulone}, {Serraller}, {Stampa}, {Tanga}, {van
  Elteren}, {van Reeven}, {Voss}, {Abbas}, {Allasia}, {Altmann}, {Anton},
  {Barache}, {Becciani}, {Berthier}, {Bianchi}, {Bombrun}, {Bouquillon},
  {Bourda}, {Bucciarelli}, {Butkevich}, {Buzzi}, {Cancelliere}, {Carlucci},
  {Charlot}, {Collins}, {Comoretto}, {Cross}, {Crosta}, {de Felice}, {Fienga},
  {Figueras}, {Fraile}, {Geyer}, {Hernandez}, {Hobbs}, {Hofmann}, {Liao},
  {Licata}, {Martino}, {McMillan}, {Michalik}, {Morbidelli}, {Parsons},
  {Pecoraro}, {Ramos-Lerate}, {Sarasso}, {Siddiqui}, {Steele},
  {Steidelm{\"u}ller}, {Taris}, {Vecchiato}, {Abreu}, {Anglada}, {Boudreault},
  {Cropper}, {Holl}, {Cheek}, {Crowley}, {Fleitas}, {Hutton}, {Osinde},
  {Rowell}, {Salguero}, {Utrilla}, {Blagorodnova}, {Soffel}, {Osorio},
  {Vicente}, {Cambras}, \& {Bernstein}}]{fabricius2016gaiadr1sourcelist}
{Fabricius}, C., {Bastian}, U., {Portell}, J., {et~al.} 2016, \aap, 595, A3

\bibitem[{{Fabricius} {et~al.}(2021){Fabricius}, {Luri}, {Arenou}, {Babusiaux},
  {Helmi}, {Muraveva}, {Reyl{\'e}}, {Spoto}, {Vallenari}, {Antoja}, {Balbinot},
  {Barache}, {Bauchet}, {Bragaglia}, {Busonero}, {Cantat-Gaudin}, {Carrasco},
  {Diakit{\'e}}, {Fabrizio}, {Figueras}, {Garcia-Gutierrez}, {Garofalo},
  {Jordi}, {Kervella}, {Khanna}, {Leclerc}, {Licata}, {Lambert}, {Marrese},
  {Masip}, {Ramos}, {Robichon}, {Robin}, {Romero-G{\'o}mez}, {Rubele}, \&
  {Weiler}}]{fabricius2021edr3cataloguevalidation}
{Fabricius}, C., {Luri}, X., {Arenou}, F., {et~al.} 2021, \aap, 649, A5

\bibitem[{{Frew} {et~al.}(2016){Frew}, {Parker}, \&
  {Boji{\v{c}}i{\'c}}}]{frewsurfacebrightness2016}
{Frew}, D.~J., {Parker}, Q.~A., \& {Boji{\v{c}}i{\'c}}, I.~S. 2016, \mnras,
  455, 1459

\bibitem[{{Gaia Collaboration} {et~al.}(2018){Gaia Collaboration}, {Brown},
  {Vallenari}, {Prusti}, {de Bruijne}, {Babusiaux}, {Bailer-Jones}, {Biermann},
  {Evans}, {Eyer}, \& et~al.}]{gaiadr2}
{Gaia Collaboration}, {Brown}, A.~G.~A., {Vallenari}, A., {et~al.} 2018, \aap,
  616, A1

\bibitem[{{Gaia Collaboration} {et~al.}(2021){Gaia Collaboration}, {Brown},
  {Vallenari}, {Prusti}, {de Bruijne}, {Babusiaux}, {Biermann}, {Creevey},
  {Evans}, {Eyer}, {Hutton}, {Jansen}, {Jordi}, {Klioner}, {Lammers},
  {Lindegren}, {Luri}, {Mignard}, {Panem}, {Pourbaix}, {Randich}, {Sartoretti},
  {Soubiran}, {Walton}, {Arenou}, {Bailer-Jones}, {Bastian}, {Cropper},
  {Drimmel}, {Katz}, {Lattanzi}, {van Leeuwen}, {Bakker}, {Cacciari},
  {Casta{\~n}eda}, {De Angeli}, {Ducourant}, {Fabricius}, {Fouesneau},
  {Fr{\'e}mat}, {Guerra}, {Guerrier}, {Guiraud}, {Jean-Antoine Piccolo},
  {Masana}, {Messineo}, {Mowlavi}, {Nicolas}, {Nienartowicz}, {Pailler},
  {Panuzzo}, {Riclet}, {Roux}, {Seabroke}, {Sordo}, {Tanga}, {Th{\'e}venin},
  {Gracia-Abril}, {Portell}, {Teyssier}, {Altmann}, {Andrae}, {Bellas-Velidis},
  {Benson}, {Berthier}, {Blomme}, {Brugaletta}, {Burgess}, {Busso}, {Carry},
  {Cellino}, {Cheek}, {Clementini}, {Damerdji}, {Davidson}, {Delchambre},
  {Dell'Oro}, {Fern{\'a}ndez-Hern{\'a}ndez}, {Galluccio}, {Garc{\'\i}a-Lario},
  {Garcia-Reinaldos}, {Gonz{\'a}lez-N{\'u}{\~n}ez}, {Gosset}, {Haigron},
  {Halbwachs}, {Hambly}, {Harrison}, {Hatzidimitriou}, {Heiter},
  {Hern{\'a}ndez}, {Hestroffer}, {Hodgkin}, {Holl}, {Jan{\ss}en}, {Jevardat de
  Fombelle}, {Jordan}, {Krone-Martins}, {Lanzafame}, {L{\"o}ffler}, {Lorca},
  {Manteiga}, {Marchal}, {Marrese}, {Moitinho}, {Mora}, {Muinonen}, {Osborne},
  {Pancino}, {Pauwels}, {Petit}, {Recio-Blanco}, {Richards}, {Riello},
  {Rimoldini}, {Robin}, {Roegiers}, {Rybizki}, {Sarro}, {Siopis}, {Smith},
  {Sozzetti}, {Ulla}, {Utrilla}, {van Leeuwen}, {van Reeven}, {Abbas}, {Abreu
  Aramburu}, {Accart}, {Aerts}, {Aguado}, {Ajaj}, {Altavilla}, {{\'A}lvarez},
  {{\'A}lvarez Cid-Fuentes}, {Alves}, {Anderson}, {Anglada Varela}, {Antoja},
  {Audard}, {Baines}, {Baker}, {Balaguer-N{\'u}{\~n}ez}, {Balbinot}, {Balog},
  {Barache}, {Barbato}, {Barros}, {Barstow}, {Bartolom{\'e}}, {Bassilana},
  {Bauchet}, {Baudesson-Stella}, {Becciani}, {Bellazzini}, {Bernet}, {Bertone},
  {Bianchi}, {Blanco-Cuaresma}, {Boch}, {Bombrun}, {Bossini}, {Bouquillon},
  {Bragaglia}, {Bramante}, {Breedt}, {Bressan}, {Brouillet}, {Bucciarelli},
  {Burlacu}, {Busonero}, {Butkevich}, {Buzzi}, {Caffau}, {Cancelliere},
  {C{\'a}novas}, {Cantat-Gaudin}, {Carballo}, {Carlucci}, {Carnerero},
  {Carrasco}, {Casamiquela}, {Castellani}, {Castro-Ginard}, {Castro Sampol},
  {Chaoul}, {Charlot}, {Chemin}, {Chiavassa}, {Cioni}, {Comoretto}, {Cooper},
  {Cornez}, {Cowell}, {Crifo}, {Crosta}, {Crowley}, {Dafonte}, {Dapergolas},
  {David}, {David}, {de Laverny}, {De Luise}, {De March}, {De Ridder}, {de
  Souza}, {de Teodoro}, {de Torres}, {del Peloso}, {del Pozo}, {Delbo},
  {Delgado}, {Delgado}, {Delisle}, {Di Matteo}, {Diakite}, {Diener},
  {Distefano}, {Dolding}, {Eappachen}, {Edvardsson}, {Enke}, {Esquej}, {Fabre},
  {Fabrizio}, {Faigler}, {Fedorets}, {Fernique}, {Fienga}, {Figueras},
  {Fouron}, {Fragkoudi}, {Fraile}, {Franke}, {Gai}, {Garabato},
  {Garcia-Gutierrez}, {Garc{\'\i}a-Torres}, {Garofalo}, {Gavras}, {Gerlach},
  {Geyer}, {Giacobbe}, {Gilmore}, {Girona}, {Giuffrida}, {Gomel}, {Gomez},
  {Gonzalez-Santamaria}, {Gonz{\'a}lez-Vidal}, {Granvik},
  {Guti{\'e}rrez-S{\'a}nchez}, {Guy}, {Hauser}, {Haywood}, {Helmi}, {Hidalgo},
  {Hilger}, {H{\l}adczuk}, {Hobbs}, {Holland}, {Huckle}, {Jasniewicz},
  {Jonker}, {Juaristi Campillo}, {Julbe}, {Karbevska}, {Kervella}, {Khanna},
  {Kochoska}, {Kontizas}, {Kordopatis}, {Korn}, {Kostrzewa-Rutkowska},
  {Kruszy{\'n}ska}, {Lambert}, {Lanza}, {Lasne}, {Le Campion}, {Le Fustec},
  {Lebreton}, {Lebzelter}, {Leccia}, {Leclerc}, {Lecoeur-Taibi}, {Liao},
  {Licata}, {Lindstr{\o}m}, {Lister}, {Livanou}, {Lobel}, {Madrero Pardo},
  {Managau}, {Mann}, {Marchant}, {Marconi}, {Marcos Santos}, {Marinoni},
  {Marocco}, {Marshall}, {Martin Polo}, {Mart{\'\i}n-Fleitas}, {Masip},
  {Massari}, {Mastrobuono-Battisti}, {Mazeh}, {McMillan}, {Messina},
  {Michalik}, {Millar}, {Mints}, {Molina}, {Molinaro}, {Moln{\'a}r},
  {Montegriffo}, {Mor}, {Morbidelli}, {Morel}, {Morris}, {Mulone}, {Munoz},
  {Muraveva}, {Murphy}, {Musella}, {Noval}, {Ord{\'e}novic}, {Orr{\`u}},
  {Osinde}, {Pagani}, {Pagano}, {Palaversa}, {Palicio}, {Panahi}, {Pawlak},
  {Pe{\~n}alosa Esteller}, {Penttil{\"a}}, {Piersimoni}, {Pineau}, {Plachy},
  {Plum}, {Poggio}, {Poretti}, {Poujoulet}, {Pr{\v{s}}a}, {Pulone}, {Racero},
  {Ragaini}, {Rainer}, {Raiteri}, {Rambaux}, {Ramos}, {Ramos-Lerate}, {Re
  Fiorentin}, {Regibo}, {Reyl{\'e}}, {Ripepi}, {Riva}, {Rixon}, {Robichon},
  {Robin}, {Roelens}, {Rohrbasser}, {Romero-G{\'o}mez}, {Rowell}, {Royer},
  {Rybicki}, {Sadowski}, {Sagrist{\`a} Sell{\'e}s}, {Sahlmann}, {Salgado},
  {Salguero}, {Samaras}, {Sanchez Gimenez}, {Sanna}, {Santove{\~n}a},
  {Sarasso}, {Schultheis}, {Sciacca}, {Segol}, {Segovia}, {S{\'e}gransan},
  {Semeux}, {Shahaf}, {Siddiqui}, {Siebert}, {Siltala}, {Slezak}, {Smart},
  {Solano}, {Solitro}, {Souami}, {Souchay}, {Spagna}, {Spoto}, {Steele},
  {Steidelm{\"u}ller}, {Stephenson}, {S{\"u}veges}, {Szabados}, {Szegedi-Elek},
  {Taris}, {Tauran}, {Taylor}, {Teixeira}, {Thuillot}, {Tonello}, {Torra},
  {Torra}, {Turon}, {Unger}, {Vaillant}, {van Dillen}, {Vanel}, {Vecchiato},
  {Viala}, {Vicente}, {Voutsinas}, {Weiler}, {Wevers}, {Wyrzykowski}, {Yoldas},
  {Yvard}, {Zhao}, {Zorec}, {Zucker}, {Zurbach}, \&
  {Zwitter}}]{gaiacollaboration2021edr3survey}
{Gaia Collaboration}, {Brown}, A.~G.~A., {Vallenari}, A., {et~al.} 2021, \aap,
  649, A1

\bibitem[{{Gaia Collaboration} {et~al.}(2016){Gaia Collaboration}, {Prusti},
  {de Bruijne}, {Brown}, {Vallenari}, {Babusiaux}, {Bailer-Jones}, {Bastian},
  {Biermann}, {Evans}, \& et~al.}]{gaiamission}
{Gaia Collaboration}, {Prusti}, T., {de Bruijne}, J.~H.~J., {et~al.} 2016,
  \aap, 595, A1

\bibitem[{{Gonz{\'a}lez-Santamar{\'\i}a}
  {et~al.}(2020){Gonz{\'a}lez-Santamar{\'\i}a}, {Manteiga}, {Manchado},
  {G{\'o}mez-Mu{\~n}oz}, {Ulla}, \& {Dafonte}}]{gonzalez2020widebinaries}
{Gonz{\'a}lez-Santamar{\'\i}a}, I., {Manteiga}, M., {Manchado}, A., {et~al.}
  2020, \aap, 644, A173

\bibitem[{{Liebert} {et~al.}(2013){Liebert}, {Bond}, {Dufour}, {Ciardullo},
  {Meakes}, {Renzini}, \& {Gianninas}}]{liebert2013hstbinaries}
{Liebert}, J., {Bond}, H.~E., {Dufour}, P., {et~al.} 2013, \apj, 769, 32

\bibitem[{Lindegren(2018)}]{gaiaruwe}
Lindegren, L. 2018, {G}AIA-C3-TN-LU-LL-124

\bibitem[{{Lindegren} {et~al.}(2021{\natexlab{a}}){Lindegren}, {Bastian},
  {Biermann}, {Bombrun}, {de Torres}, {Gerlach}, {Geyer}, {Hern{\'a}ndez},
  {Hilger}, {Hobbs}, {Klioner}, {Lammers}, {McMillan}, {Ramos-Lerate},
  {Steidelm{\"u}ller}, {Stephenson}, \& {van
  Leeuwen}}]{lindegren2021parallaxzeropoint}
{Lindegren}, L., {Bastian}, U., {Biermann}, M., {et~al.} 2021{\natexlab{a}},
  \aap, 649, A4

\bibitem[{{Lindegren} {et~al.}(2021{\natexlab{b}}){Lindegren}, {Klioner},
  {Hern{\'a}ndez}, {Bombrun}, {Ramos-Lerate}, {Steidelm{\"u}ller}, {Bastian},
  {Biermann}, {de Torres}, {Gerlach}, {Geyer}, {Hilger}, {Hobbs}, {Lammers},
  {McMillan}, {Stephenson}, {Casta{\~n}eda}, {Davidson}, {Fabricius},
  {Gracia-Abril}, {Portell}, {Rowell}, {Teyssier}, {Torra}, {Bartolom{\'e}},
  {Clotet}, {Garralda}, {Gonz{\'a}lez-Vidal}, {Torra}, {Abbas}, {Altmann},
  {Anglada Varela}, {Balaguer-N{\'u}{\~n}ez}, {Balog}, {Barache}, {Becciani},
  {Bernet}, {Bertone}, {Bianchi}, {Bouquillon}, {Brown}, {Bucciarelli},
  {Busonero}, {Butkevich}, {Buzzi}, {Cancelliere}, {Carlucci}, {Charlot},
  {Cioni}, {Crosta}, {Crowley}, {del Peloso}, {del Pozo}, {Drimmel}, {Esquej},
  {Fienga}, {Fraile}, {Gai}, {Garcia-Reinaldos}, {Guerra}, {Hambly}, {Hauser},
  {Jan{\ss}en}, {Jordan}, {Kostrzewa-Rutkowska}, {Lattanzi}, {Liao}, {Licata},
  {Lister}, {L{\"o}ffler}, {Marchant}, {Masip}, {Mignard}, {Mints}, {Molina},
  {Mora}, {Morbidelli}, {Murphy}, {Pagani}, {Panuzzo}, {Pe{\~n}alosa Esteller},
  {Poggio}, {Re Fiorentin}, {Riva}, {Sagrist{\`a} Sell{\'e}s}, {Sanchez
  Gimenez}, {Sarasso}, {Sciacca}, {Siddiqui}, {Smart}, {Souami}, {Spagna},
  {Steele}, {Taris}, {Utrilla}, {van Reeven}, \&
  {Vecchiato}}]{lindegren2021edr3astrometry}
{Lindegren}, L., {Klioner}, S.~A., {Hern{\'a}ndez}, J., {et~al.}
  2021{\natexlab{b}}, \aap, 649, A2

\bibitem[{{Luri} {et~al.}(2018){Luri}, {Brown}, {Sarro}, {Arenou},
  {Bailer-Jones}, {Castro-Ginard}, {de Bruijne}, {Prusti}, {Babusiaux}, \&
  {Delgado}}]{xlurigaiaparallaxes}
{Luri}, X., {Brown}, A.~G.~A., {Sarro}, L.~M., {et~al.} 2018, \aap, 616, A9

\bibitem[{Parker {et~al.}(2006)Parker, Acker, Frew, Hartley, Peyaud,
  Ochsenbein, Phillipps, Russeil, Beaulieu, Cohen, Köppen, Miszalski, Morgan,
  Morris, Pierce, \& Vaughan}]{mashpn}
Parker, Q.~A., Acker, A., Frew, D.~J., {et~al.} 2006, Monthly Notices of the
  Royal Astronomical Society, 373, 79

\bibitem[{{Parker} {et~al.}(2016){Parker}, {Boji{\v{c}}i{\'c}}, \&
  {Frew}}]{hashpn}
{Parker}, Q.~A., {Boji{\v{c}}i{\'c}}, I.~S., \& {Frew}, D.~J. 2016, in Journal
  of Physics Conference Series, Vol. 728, Journal of Physics Conference Series,
  032008

\bibitem[{{Price-Whelan} {et~al.}(2018){Price-Whelan}, {Sip{\H{o}}cz},
  {G{\"u}nther}, {Lim}, {Crawford}, {Conseil}, {Shupe}, {Craig}, {Dencheva},
  {Ginsburg}, {VanderPlas}, {Bradley}, {P{\'e}rez-Su{\'a}rez}, {de Val-Borro},
  {Paper Contributors}, {Aldcroft}, {Cruz}, {Robitaille}, {Tollerud},
  {Coordination Committee}, {Ardelean}, {Babej}, {Bach}, {Bachetti}, {Bakanov},
  {Bamford}, {Barentsen}, {Barmby}, {Baumbach}, {Berry}, {Biscani}, {Boquien},
  {Bostroem}, {Bouma}, {Brammer}, {Bray}, {Breytenbach}, {Buddelmeijer},
  {Burke}, {Calderone}, {Cano Rodr{\'\i}guez}, {Cara}, {Cardoso}, {Cheedella},
  {Copin}, {Corrales}, {Crichton}, {D{\textquoteright}Avella}, {Deil},
  {Depagne}, {Dietrich}, {Donath}, {Droettboom}, {Earl}, {Erben}, {Fabbro},
  {Ferreira}, {Finethy}, {Fox}, {Garrison}, {Gibbons}, {Goldstein}, {Gommers},
  {Greco}, {Greenfield}, {Groener}, {Grollier}, {Hagen}, {Hirst}, {Homeier},
  {Horton}, {Hosseinzadeh}, {Hu}, {Hunkeler}, {Ivezi{\'c}}, {Jain}, {Jenness},
  {Kanarek}, {Kendrew}, {Kern}, {Kerzendorf}, {Khvalko}, {King}, {Kirkby},
  {Kulkarni}, {Kumar}, {Lee}, {Lenz}, {Littlefair}, {Ma}, {Macleod},
  {Mastropietro}, {McCully}, {Montagnac}, {Morris}, {Mueller}, {Mumford},
  {Muna}, {Murphy}, {Nelson}, {Nguyen}, {Ninan}, {N{\"o}the}, {Ogaz}, {Oh},
  {Parejko}, {Parley}, {Pascual}, {Patil}, {Patil}, {Plunkett}, {Prochaska},
  {Rastogi}, {Reddy Janga}, {Sabater}, {Sakurikar}, {Seifert}, {Sherbert},
  {Sherwood-Taylor}, {Shih}, {Sick}, {Silbiger}, {Singanamalla}, {Singer},
  {Sladen}, {Sooley}, {Sornarajah}, {Streicher}, {Teuben}, {Thomas},
  {Tremblay}, {Turner}, {Terr{\'o}n}, {van Kerkwijk}, {de la Vega}, {Watkins},
  {Weaver}, {Whitmore}, {Woillez}, {Zabalza}, \& {Contributors}}]{astropy:2018}
{Price-Whelan}, A.~M., {Sip{\H{o}}cz}, B.~M., {G{\"u}nther}, H.~M., {et~al.}
  2018, \aj, 156, 123

\bibitem[{{Riello} {et~al.}(2021){Riello}, {De Angeli}, {Evans}, {Montegriffo},
  {Carrasco}, {Busso}, {Palaversa}, {Burgess}, {Diener}, {Davidson}, {Rowell},
  {Fabricius}, {Jordi}, {Bellazzini}, {Pancino}, {Harrison}, {Cacciari}, {van
  Leeuwen}, {Hambly}, {Hodgkin}, {Osborne}, {Altavilla}, {Barstow}, {Brown},
  {Castellani}, {Cowell}, {De Luise}, {Gilmore}, {Giuffrida}, {Hidalgo},
  {Holland}, {Marinoni}, {Pagani}, {Piersimoni}, {Pulone}, {Ragaini}, {Rainer},
  {Richards}, {Sanna}, {Walton}, {Weiler}, \&
  {Yoldas}}]{riello2021edr3photometry}
{Riello}, M., {De Angeli}, F., {Evans}, D.~W., {et~al.} 2021, \aap, 649, A3

\bibitem[{{Sch{\"o}nberner} {et~al.}(2018){Sch{\"o}nberner}, {Balick}, \&
  {Jacob}}]{expansion2018}
{Sch{\"o}nberner}, D., {Balick}, B., \& {Jacob}, R. 2018, \aap, 609, A126

\bibitem[{{Smith}(2015)}]{smith2015}
{Smith}, H. 2015, \mnras, 449, 2980

\bibitem[{{Sutherland} \& {Saunders}(1992)}]{sutherlandsaunders1992}
{Sutherland}, W. \& {Saunders}, W. 1992, \mnras, 259, 413

\bibitem[{{Torra} {et~al.}(2021){Torra}, {Casta{\~n}eda}, {Fabricius},
  {Lindegren}, {Clotet}, {Gonz{\'a}lez-Vidal}, {Bartolom{\'e}}, {Bastian},
  {Bernet}, {Biermann}, {Garralda}, {G{\'u}rpide}, {Lammers}, {Portell}, \&
  {Torra}}]{torra2021gaiaedr3sourcelist}
{Torra}, F., {Casta{\~n}eda}, J., {Fabricius}, C., {et~al.} 2021, \aap, 649,
  A10

\bibitem[{{Weidmann} \& {Gamen}(2011)}]{weidmannCSPNe}
{Weidmann}, W.~A. \& {Gamen}, R. 2011, \aap, 526, A6

\end{thebibliography}

\begin{appendix}
\section{Best matches table}
\begin{onecolumn}
\begin{tiny}
\begin{longtable}{lp{0.15\textwidth}lccccccc}
\caption[Long table]{\label{table:catalogue}Best matches.\protect\footnotemark}\\
\hline\hline
PN G & PN name & \textit{Gaia} source ID & reliability & RA & Dec & G & $\omega$\footnotemark & $\sigma_\omega$\footnotemark & distance\footnotemark \\
 &  &  &  & (deg) & (deg) & (mag) & (mas) & (mas) & (kpc) \\
\hline
\endfirsthead
\caption{continued.}\\
\hline\hline
PN G & PN name & \textit{Gaia} source ID & reliability & RA & Dec & G & $\omega$ & $\sigma_\omega$ & distance \\
 &  &  &  & (deg) & (deg) & (mag) & (mas) & (mas) & (kpc) \\
\hline
\endhead
\hline
\endfoot
\addtocounter{footnote}{-4}\stepcounter{footnote}\footnotetext{Only the most important columns are shown due to space considerations. The complete set of columns will be published on CDS, and missing columns may also be obtained by joining the table against \textit{Gaia} EDR3 and HASH.}\stepcounter{footnote}\footnotetext{Parallax, not corrected for zero point.}\stepcounter{footnote}\footnotetext{Parallax errror.}\stepcounter{footnote}\footnotetext{Median value of the posterior; upper and lower bounds are 84th and 16th percentiles. See Sect. \ref{sect:distancescombined}.}000.0$+$02.0 & K 6-4 & 4060525376634471808 & 0.65 & $264.42892$ & $-27.81947$ & 19.5 & $\ldots$ & $\ldots$ & $\ldots$ \\
000.0$-$01.0 & JaSt 69 & 4057224264723323264 & 0.26 & $267.54157$ & $-29.31827$ & 19.7 & $-0.817$ & $0.396$ & $\ldots$ \\
000.0$-$06.8 & H 1-62 & 4045771305065496832 & 0.98 & $273.32482$ & $-32.32860$ & 14.4 & $0.047$ & $0.034$ & $8.95^{+1.73}_{-1.47}$ \\
000.1$+$04.3 & H 1-16 & 4109679250049965312 & 0.98 & $262.34731$ & $-26.43461$ & 18.7 & $\ldots$ & $\ldots$ & $\ldots$ \\
000.1$+$17.2 & PC 12 & 4130784921205604736 & 1.00 & $250.97408$ & $-18.95333$ & 15.2 & $0.069$ & $0.066$ & $9.86^{+3.01}_{-2.58}$ \\
000.1$-$05.6 & H 2-40 & 4049045783774253696 & 0.57 & $272.12813$ & $-31.60981$ & 18.5 & $0.340$ & $0.268$ & $5.94^{+2.01}_{-1.94}$ \\
000.1$-$08.0 & SB 1 & 4045583322840954496 & 1.00 & $274.70193$ & $-32.79860$ & 20.0 & $-0.228$ & $0.616$ & $\ldots$ \\
000.2$+$01.4 & PBOZ 26 & 4060650446035698944 & 0.75 & $265.16073$ & $-28.01722$ & 21.1 & $\ldots$ & $\ldots$ & $\ldots$ \\
\hline
\end{longtable}
\end{tiny}
\end{onecolumn}
\end{appendix}

\end{document}